\documentclass[twocolumn,preprintnumbers,nofootinbib,prd,superscriptaddress,aps]{revtex4-1}

\usepackage[utf8]{inputenc}
\usepackage{graphicx,amssymb,amsmath,amsthm,amsfonts,epstopdf,epsfig,epsf,times}
\usepackage[linktocpage]{hyperref}
\usepackage[usenames]{color}
\usepackage{epstopdf}

\usepackage{aas_macros}
\usepackage{bm}
\usepackage{dcolumn}
\usepackage{latexsym}
\usepackage{rotating}
\usepackage{hyperref}
\usepackage{color}
\usepackage{longtable}
\usepackage{enumerate}
\usepackage{tensor}
\usepackage{stmaryrd}

\usepackage{float}
\usepackage[caption = false]{subfig}

\usepackage{mathtools}
\usepackage{url}
\definecolor{coolblack}{rgb}{0.0, 0.18, 0.39}
\definecolor{darkred}{rgb}{0.5,0,0}
\definecolor{darkgreen}{rgb}{0,0.5,0}
\definecolor{darkblue}{rgb}{0,0,0.5}
\definecolor{lapislazuli}{rgb}{0.15, 0.38, 0.61}
\definecolor{venetianred}{rgb}{0.78, 0.03, 0.08}
\definecolor{bleudefrance}{rgb}{0.19, 0.55, 0.91}
\definecolor{dogwoodrose}{rgb}{0.84, 0.09, 0.41}
\definecolor{dogwoodrose}{rgb}{0.84, 0.09, 0.41}
\hypersetup{colorlinks=true, citecolor=darkblue, linkcolor=darkblue, 
urlcolor = darkblue}

\graphicspath{{pictures/}} %setting the picture directory

\definecolor{olive}{rgb}{0.5, 0.5, 0.0}

\renewcommand{\vec}[1]{\boldsymbol{#1}}

\newcommand{\ben}{\begin{enumerate}}
\newcommand{\een}{\end{enumerate}}

\def\be{\begin{equation}}
\def\ee{\end{equation}}
\newcommand{\beq}{\begin{eqnarray}}
\newcommand{\eeq}{\end{eqnarray}} 
\newcommand{\ba}{\begin{align}}
\newcommand{\ea}{\end{align}}

\def\be{\begin{equation}}
\def\ee{\end{equation}}
	
\newcommand{\bea}{\begin{eqnarray}}
\newcommand{\eea}{\end{eqnarray}}

\begin{document}\title {\large How do scalar-field dark matter haloes react to orbiting bodies?}

\author{Miguel C. Ferreira}
\affiliation{CENTRA, Departamento de F\'{\i}sica, Instituto Superior T\'ecnico -- IST, Universidade de Lisboa -- UL,
Avenida Rovisco Pais 1, 1049 Lisboa, Portugal}

\begin{abstract}
Low-energy, self-gravitating solutions of a scalar field coupled to gravity, described by the Schrodinger-Poisson system, are good candidates for realistic astrophysical structures, being particularly suited to describe dark matter halos. In this work we study the scenario in which one of these structures is gravitationally perturbed by a point-like mass. We analyse the effects that the body has on the distribution of the scalar field and how it backreacts on the body's motion. We show that an initially static, spherical structure can develop rotating non-spherical clumps, the amplitude and the velocity of which are directly related to the mass of the orbiting particle. We also study the dissipation mechanisms involved in the transit of the point-like particle across the scalar field structure and we observe that the force responsible for the dissipation scales as the square of the mass of the particle.
\end{abstract}

\date{\today}

\maketitle

%TABLE OF CONTENTS
\tableofcontents

%%%%%%%%%%%%%%%%%%%%%%%%%%%%%%%%%%%%%%%%%%%%%%
\section{Introduction}\label{sec:intro}
%%%%%%%%%%%%%%%%%%%%%%%%%%%%%%%%%%%%%%%%%%%%%%

Scalar fields are the simplest objects one can have in a field theory. Their simplicity is the key for their versatility which puts them in the center of the most important debates of modern physics. Scalar fields are present in many areas of study, either as by-products in low energy limits of string theory~\cite{Svrcek:2006yi, Arvanitaki:2009fg} or as postulated ingredients of models that try to explain poorly-understood phenomena, such as inflation~\cite{Nanopoulos:1983up,Baer:2014eja}, quintessence~\cite{Ratra:1987rm, Caldwell:1997ii} or the QCD's strong CP problem~\cite{Peccei:1977hh}.

The plethora of scalar fields that appear in these scenarios are predicted in such a way that they interact very weakly with baryonic matter, making them hard to reveal in particle detectors. In any case, all of these fields have to interact gravitationally and it is at astrophysical scales that their gravitational influence may become big enough for their presence to be detected~\cite{Kesden:2004qx,Macedo:2013qea,Khmelnitsky:2013lxt,Eilers:2013lla,Macedo:2013jja,Blas:2016ddr,Ferreira:2017pth,Boskovic:2018rub,Boskovic:2018lkj}. 

One case of interest happens when scalar fields develop structures -- dubbed ``scalar clouds'' -- around black holes (BHs) as a result of the activation of a superradiance mechanism~\cite{Brito:2015oca}. These clouds can have a strong impact on the gravitational wave-driven dynamics of BHs, creating a fertile ground for astrophysical tests~\cite{Brito:2014wla, Arvanitaki:2014wva, Herdeiro:2015waa, Cunha:2015yba, Ferreira:2017pth,Brito:2017zvb,Brito:2017wnc,Ikeda:2018nhb}. Another astrophysically relevant scenario involving scalar fields is the formation of self-gravitating structures known as boson stars or oscillatons, for complex and real scalars, respectively~\cite{Kaup:1968zz,Ruffini:1969qy,Brito:2015yga,Brito:2015pxa,Seidel:1991zh,Liebling:2012fv}. These scalar structures are solutions of the Einstein-Klein-Gordon system of equations and their characteristics can vary substantially depending on the scalar field potential that is considered. Indeed, different potentials give rise to structures that can either resemble BHs~\cite{Guzman:2009zz,Palenzuela:2017kcg} or big galactic haloes \cite{UrenaLopez:2010ur, Lee:1995af}. The latter case has motivated the use of scalar fields as a viable model to explain the phenomena atributed to dark matter~\cite{Mielke:2002bp, Hui:2016ltb, Hu:2000ke}. The formation of boson stars and oscillatons has been studied \cite{Seidel:1993zk, Madsen:1990gg, 1985MNRAS.215..575K, Guzman:2006yc,Okawa:2013jba, Hidalgo:2017dfp,Veltmaat:2018dfz}) as well as their stability \cite{Jetzer:1988vr, Gleiser:1988rq, Lee:1988av, Seidel:1990jh, Balakrishna:1997ej, Guzman:2004wj,Grandclement:2011wz,Alcubierre:2003sx,Brito:2017zvb,Brito:2017wnc}. They can exist in a ground state (a 0-node solution) or they can be formed in an excited state (a $n$-node solution, $n \neq 0$) which either decays to the ground state, collapses to a BH or dissipates away \cite{Guzman:2006yc,Guzman:2004wj,Balakrishna:2007mr,Guzman:2004jw}.

Infering the existence of scalar fields from its self-gravitating structures requires that one knows what observational imprints those structures may leave. This line of research is a very active one, some exemples of which are the analysis of geodesics around these structures~\cite{Eilers:2013lla,Macedo:2013jja,Kesden:2004qx,Macedo:2013qea,Brihaye:2014gua,Grandclement:2014msa,2006GReGr..38..633B,Boskovic:2018rub}, the study of their gravitational lensing effects~\cite{Schunck:2006rk,Dabrowski:1998ac}, their emission of gravitational radiation \cite{Balakrishna:2006ru,Ferrell:1989kz,Kesden:2004qx}, its behavior when involved in collisions with other structures~\cite{Eby:2017xaw,Palenzuela:2006wp,Palenzuela:2017kcg,Clough:2018exo} and their effects on the pulsar timing signal measurements \cite{Khmelnitsky:2013lxt,Blas:2016ddr}.

In the context of scalar field dark matter haloes~\cite{Mielke:2002bp,Hui:2016ltb,Hu:2000ke}, only the self-gravitating structures which are characterized by low compacteness and large radii are of interest. Boson stars and oscillatons which comply with these requirements can be studied within a Newtonian setup. In fact, when the magnitude of the scalar field is very small, the solutions of the Einstein-Klein-Gordon system have extremely large spatial extent and the underlying spacetime geometry is very close to Minkowski. It was shown \cite{Ferrell:1989kz,Guzman:2004wj,Boskovic:2018rub} that to study this regime, it is sufficient to work with the simpler Schrodinger-Poisson (SP) system instead of the Einstein-Klein-Gordon one. Moreover, the SP system describes the dynamics of both complex and real scalar fields \cite{Guzman:2004wj} in the low-energy limit.

Since the low-energy scalar configurations in the ground state are stable, one can assume that they may be astrophysically relevant as components of bounded gravitational systems. We are particularly interested in a two-body system in which one of the bodies is described by a stable scalar field configuration and the other is pointlike. In this scenario, the scalar field configuration is tidally deformed due to the presence of the point-particle and that deformation affects the evolution of the whole system. We developed a numerical code to study this two-body scenario and we are able to describe the tidal deformations of the scalar field density as well as the gravitational friction force that the point-particle feels as it traverses the scalar configuration.

We use Planck units -- $c = \hbar = G = 1$ -- unless otherwise stated.

%%%%%%%%%%%%%%%%%%%%%%%%%%%%%%%%%%%%%%%%%%%%%%%%%%%%%%
\section{Framework}
\label{sec:describing-the-system}
%%%%%%%%%%%%%%%%%%%%%%%%%%%%%%%%%%%%%%%%%%%%%%%%%%%%%%
We consider the scenario in which a low-energy, stable, scalar field configuration of the Einstein-Klein Gordon system is perturbed by a pointlike mass. This two-component system will be evolved separately and the only interaction between the components is gravitational. The time evolution of the low-energy scalar struture will be described by the SP equations, which will contain the effect of the presence of the point-particle, whereas the movement of the point-particle will be rendered from the gravitational potential of the scalar field configuration by the laws of Newtonian mechanics.

%%%%%%%%%%%%%%%%%%%%%%%%%%%%%%%%%%%%%%%%%%%%%%%%%%%%%%%%%%%%%%%%%%%%%%%%%%%%%%
\subsection{Low energy scalar field structures\label{subsec:SP-stationary}}
%%%%%%%%%%%%%%%%%%%%%%%%%%%%%%%%%%%%%%%%%%%%%%%%%%%%%%%%%%%%%%%%%%%%%%%%%%%%%%

A complex scalar field of mass $m_s$, minimally coupled to gravity, is described by the action
\begin{equation}
  S = \int d^4 x \sqrt{- g} \left( \frac{R}{16\pi} - \frac{1}{2} g^{\alpha \beta} \partial_{\alpha} \Phi^* \partial_{\beta} \Phi - \frac{1}{2} m_s^2 \Phi \Phi^*\right),
\end{equation}
% 
% ($\mu = m_sc/\hbar$ in \mf{``normal units''}).
Minimizing this action is equivalent to solving the Einstein-Klein-Gordon system
\beq
&&R_{\alpha \beta} - \frac{R}{2} g_{\alpha \beta} = 8\pi T_{\alpha \beta}\,,\\
&&T_{\alpha \beta} = \frac{1}{2}\left(\Phi_{,\alpha} \Phi_{,\beta}^* + \Phi_{,\alpha}^* \Phi_{,\beta}\right) - \frac{g_{\alpha \beta}}{2} \left(\Phi^{,\gamma} \Phi_{,\sigma}^* + m_s^2 \Phi \Phi^*\right)\,,\\
&&\frac{1}{\sqrt{-g}} \partial_{\alpha}\left( \sqrt{-g} g^{\alpha \beta} \partial_{\beta} \Phi \right) =m_s^2 \Phi\,,\label{eq:EKG-system}
\eeq
where $R_{\alpha \beta}$ and $R$ represent the Ricci tensor and Ricci scalar, respectively, $g_{\alpha\beta}$ is the metric tensor and $g$ is its determinant. We are interested in the low-energy limit, i.e., in situations in which the scalar field energy is given almost entirely by its rest-energy and the space-time metric deviates slightly from the Minkowski metric. In the harmonic gauge, $\partial_{\nu} \left( \sqrt{-g} g^{\mu \nu} \right) = 0$, the metric tensor can be written as
\be
g_{00}=-1+2 U(r)\,,\qquad g_{jk}=\left[1+2 U(r)\right]\delta_{jk}\,,\label{eq:metric-ansatz}
\ee
and $\quad g_{0j} = 0$. The scalar field has an harmonic time dependence
\be 
\Phi \sim \exp(-\mathrm{i} m_s t) \chi(t,r)\,, \label{eq:low-energy-field}
\ee
with $\chi(t,r)$ a slowly varying function of time. Notice that the spatial dependence is solely radial, since we're assuming spherical symmetry. Substituting Eqs.~\eqref{eq:metric-ansatz} and \eqref{eq:low-energy-field} in the Einstein-Klein-Gordon system and focusing only on the leading order terms, one obtains the Schrodinger-Poisson (SP) system
\beq
\mathrm{i}\partial_t \chi &=& - \frac{1}{2 m_s} \nabla^2 \chi + m_s U \chi\,, \nonumber \\
\nabla^2 U &=& 4\pi m_s^2 |\chi|^2\,.\label{eq:SP-unscalled}
\eeq
Making the following re-scaling
\be
t \rightarrow m_s t, \quad r \rightarrow m_s r, \quad \chi \rightarrow \frac{1}{\sqrt{4 \pi}} \psi\,,\label{eq:variable-normalization}
\ee
the previous equations can be written as
\beq
\mathrm{i} \partial_{t} \psi &=& -\frac{1}{2}\nabla^2 \psi + U \psi,\nonumber\\
\nabla^2 U &=& \psi \psi^*.
\eeq
It is known~\cite{Guzman:2004wj} that a transformation of the form 
\begin{equation}
\label{eq:scaling}
(t ,r, U, \psi) \rightarrow (\lambda^{-2} \hat{t}, \lambda^{-1} \hat{r}, \lambda^2 \hat{U}, \lambda^2\hat{\psi}),
\end{equation}
leaves the SP system of equations unchanged. Using this property, one can normalize the system by working only with the ``hat-variables'', meaning that the order of magnitude of all the quantities involved in this problem is hidden in the parameter $\lambda$. Notice that Ref.~\cite{UrenaLopez:2002gx} fixes the value of $\lambda$ for which the behavior described by the SP sytem coincides with the behavior of the Einstein-Klein-Gordon system (i.e. the scale of the low-energy limit)
\begin{equation}
\lambda^2 < 10^{-3}\,.
\end{equation}
This is the limit of validity of all the statements that stem from the analysis of the SP system: these calculations only cover scalar fields whose magnitude is compatible with the previous limit. So, having established the limits of validity of our working system, we will, from now on and unless otherwise states, work in terms of the ``hat-quantities'' of Eq.~\eqref{eq:scaling}.

\begin{figure}
\centering
\includegraphics[width=\columnwidth]{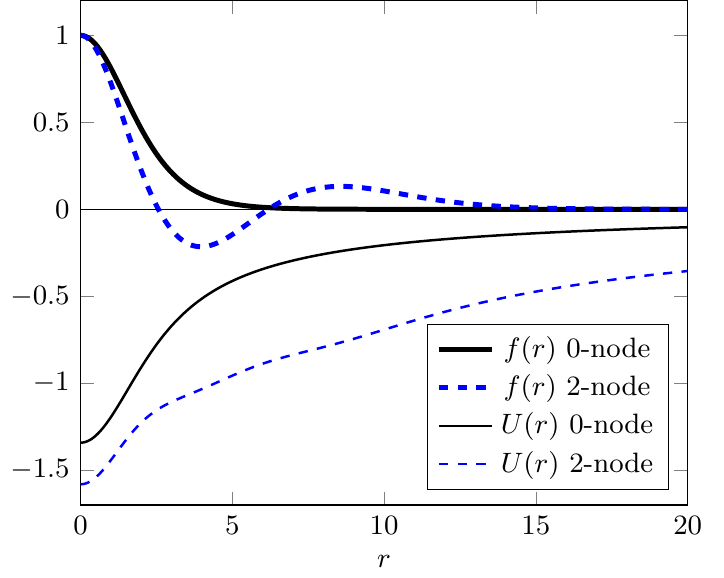}
\caption{Representing two stationary profiles of the SP system with the the normalized ``hat quantities'' of Eq.~\eqref{eq:scaling}. In black we see the 0-node configuration of the scalar field (thick line) and of the gravitational potential (thin line); in dashed blue it is represented the 2-node configuration in the same way. Notice that the gravitational potential of the 2-node configuration is deeper than the one of the 0-node configuration.}
\label{fig:stationary-profiles}
\end{figure}
Stationary solutions for the SP system have been found \cite{Guzman:2004wj} and their stability was studied. These solutions are spherically symmetric and are classified by the number of nodes they present, i.e., by the number of times that the scalar field function changes sign. It was found that only the fundamental (0-node) solutions are stable; all the other stationary solutions decay into nodeless configurations by losing mass through scalar radiation~\cite{Guzman:2004wj}. To obtain these stationary solutions, we consider a scalar field of the form
\begin{equation}
\psi = \exp(-\mathrm{i} \gamma t) f(r),
\end{equation}
i.e. $\gamma$ is the difference between the total energy of the field and its rest energy (notice that $\psi$ corresponds to the slow-varying part of the scalar field, the total scalar field function being $\Phi \sim \exp(-\mathrm{i}m_s t) \exp(-\mathrm{i} \gamma t) f(r)$; thus, its energy is $E = m_s + \gamma$)\footnote{Since we are in a low-energy limit, this difference is very small, however, since we are working with the ``hat-quantities'' (see Eq.~\eqref{eq:scaling}), the value of $\gamma$ that we present in the equations is related with that normalized system of coordinates; to convert it back to Planck units, one has to multiply it by $\lambda^2$, which will consistently make it small} and since stationary configurations are bound states of the system, one expects\footnote{Using the non-scaled variables of Eq.~\eqref{eq:SP-unscalled}, one can see that since $\frac{\gamma}{m_s} = \frac{\omega}{m_s} - 1$, and because the bound state condition is $\frac{\omega}{m_s} < 1$, a bound state must have $\frac{\gamma}{m_s} <0$.} that $\gamma <0$. By substituting the previous ansatz on the SP system, one obtains
\beq
&&f''(r) + \frac{2}{r} f'(r) + 2 (\gamma - U(r))f(r) = 0\,, \label{eq:stationary-solutions-equation}\\
&&U''(r) + \frac{2}{r} U'(r) = f(r) f^*(r)\,. \nonumber
\eeq
These equations are used to find the profiles $f(r)$ of the stationary configurations of the scalar field. To do it, one has to impose boundary conditions that come from two reasonable physical requirements: the profile has to be regular and finite. Regularity is enforced by demanding that $f'(r)$ and $U'(r)$ are zero in the origin; finiteness is then guaranteed by insisting that $\lim_{r\rightarrow \infty} f(r) = 0$. Moreover, one must demand that the resulting gravitational potential $U(r)$, when measured at infinity, describes, as it should, the effect of the total mass of the scalar field configuration. This quantity is calculated with the volume integral
\begin{equation}
M_{\rm DM} = \int \rho(t,r) dV\,,
\end{equation}
where $\rho = m_s^2 \Phi \Phi^*$ is the leading order term of the the weak-field limit of the $00$ component of the scalar field energy-momentum tensor. Writing the previous integral in terms of the hat-quantities of Eq.~\eqref{eq:scaling}, we obtain $M_{\rm DM} = M_f/m_s$ where
\begin{equation}
\label{eq:stationary-mass}
M_f = \int \rho_f(t,r) dV = \frac{1}{4\pi} \int \psi \psi^* dV\,,
\end{equation}
which in the case of the stationary configuration, can be simplified to
\begin{equation}
  M_f =  \frac{1}{4\pi} \int_0^{\infty}\psi^* \psi  dV =  \int_0^{\infty}  f(r) f^*(r) r^2  dr\,.
\end{equation}
Having calculated the mass, one can write the boundary behavior for the gravitational potential in the hat-quantities as
\begin{equation}
\lim_{r\rightarrow \infty} U(r) = -\frac{M_f}{r}.
\end{equation}
Imposing these conditions along with the value for the scalar field at the origin (since we are working with the normalized ``hat quantities'' it is enough to consider $f(0) = 1$), one obtains profiles $f(r)$ which can be characterized by the number of nodes. To each of the profiles corresponds a unique value of the quantities $\gamma$ and $U(r=0)$. In Fig.~\ref{fig:stationary-profiles} we show two stationary configurations and their respective potential.
Since we are only interested in the 0-node, ground state solutions, we quote here only its characteristic values,
\beq
\gamma&=& -0.6922, \quad U(0) = -1.3418,\nonumber\\
M_f &=& 2.0622,\quad R_{99} =4.8228 \,,\label{eq:properties-equilibrium}
\eeq
where $R_{99}$ is the radial position up to which $99\%$ of the mass of the scalar configuration is contained.

%%%%%%%%%%%%%%%%%%%%%%%%%%%%%%%%%%%%%%%%%%%%%%%%%%%%%%%%%%%%%%%%%%%%%
\subsection{The point-like particle}
%%%%%%%%%%%%%%%%%%%%%%%%%%%%%%%%%%%%%%%%%%%%%%%%%%%%%%%%%%%%%%%%%%%%%
We now want to understand how an orbiting mass $M_p$ disturbs, dynamically, the previous self-gravitating massive scalar structure.
We model the orbiting mass as pointlike~\footnote{Due to the nonlinear nature of Einstein's field equations, problems arise in the definition of point particles in such context. However, we will always be working in the Newtonian limit where such idealization is acceptable.}
and use its energy-momentum tensor in the Einstein-Klein-Gordon system of Eqs.~\eqref{eq:EKG-system} (see section 6.5 of Ref.~\cite{straumann2012general})
\begin{equation}
  \label{eq:em-point-particle-integral}
T^{\mu\nu}_P = \frac{1}{\sqrt{-g}} M_p \int \frac{d x_P^{\mu}}{d \tau} \frac{d x_P^{\nu}}{d \tau} \delta^{(4)}(x^{\alpha} - x_P^{\alpha}(\tau)) \mathrm{d}\tau\,,
\end{equation}
where $x^{\alpha}_P$ are the spacetime coordinates of the point-particle, $\tau$ is its proper-time and $\delta^{(4)}$ is the Dirac-delta. A low-energy analysis of the Einstein-Klein-Gordon system with the pointlike mass included yields (see Appendix~\ref{app:low-energy})
\beq
\mathrm{i}\partial_t \psi &=& - \frac{1}{2} \nabla^2 \psi + U \psi\,, \nonumber \\
\nabla^2 U &=&  \psi \psi^* + P(\vec{x},\vec{x}_P)\,,\label{eq:SP-scaled-dirac}
\eeq
with
\begin{equation}
  P(\vec{x},\vec{x}_P) = \frac{4 \pi M_P}{r^2}  \delta(r - r_P) \delta(\cos\theta - \cos\theta_P) \delta(\phi - \phi_P)\,,\label{stress_pp}
\end{equation}
where $r_P$, $\theta_P$ and $\phi_P$ correspond to the spherical coordinates indicating the position of the point-like particle. Notice that since we are using the ``hat quantities'' of Eq.~\eqref{eq:scaling}, the value of the mass of the orbiting particle, $M_p$, when converted to Planck units, must also be multiplied by $m_s$. Using the spherical harmonics closure relation~\cite{arfken2011mathematical}
\begin{equation}
\delta(\cos\theta - \cos\theta_P) \delta(\phi-\phi_P) = \sum_{\ell m} Y_{\ell m}^*(\theta_P,\phi_P) Y_{\ell m} (\theta,\phi)\,,\nonumber
\end{equation}
Eq.~\eqref{stress_pp} can be re-written as a sum of spherical harmonics
\begin{equation}
  P(\vec{x},\vec{x}_P) = p_A(r,\vec{x}_P) + \sum_{\ell,m} p_{\ell, m}(r,\vec{x}_P) Y_{\ell m} (\theta,\phi)\,, \label{eq:dirac-delta-particle}
\end{equation}
where
\beq
p_A(r,\vec{x}_P) &=& \frac{M_P \delta(r - r_P)}{r^2},\nonumber\\
p_{\ell,m}(r,\vec{x}_P) &=& \frac{4 \pi M_P \delta(r - r_P)}{r^2} Y_{\ell m}^*(\theta_P,\phi_P),
\eeq
This expansion motivates us to write the other quantities involved in the problem in a similar way, i.e.
\beq
\psi(r,\theta,\phi) & =& \varphi_A(r) + \sum_{\ell,m} \varphi_{\ell, m}(r) Y_{\ell m} (\theta,\phi), \label{eq:field-components}\\
U(r,\theta,\phi) & =& V_A(r)+ \sum_{\ell,m} V_{\ell, m}(r) Y_{\ell m} (\theta,\phi)\label{eq:potential-components}.
\eeq
%

%%%%%%%%%%%%%%%%%%%%%%%%%%%%%%%%%%%%%%%%%
\subsection{The evolution equations}
%%%%%%%%%%%%%%%%%%%%%%%%%%%%%%%%%%%%%%%%%

When the pointlike particle is put in orbit, both the scalar field and the gravitational potential will, in general, develop a multipolar strucuture which will translate in the development of non-trivial profiles for the $(\ell, m)$ components of the expansions of Eqs.~\eqref{eq:field-components} and \eqref{eq:potential-components}. This development will be coordinated by the Schrodinger-Poisson system of equations, where we introduce the aforementioned expressions for the scalar field and gravitational potential, which are then projected in each $(\ell, m)$ component. By doing that, we obtain two equations for each mode, one for $\varphi_{\ell,m}$, coming from the projection of the Schrodinger equation, and one for $V_{\ell,m}$ coming from the projection of the Poisson equation. Particularly, in order to harvest the zeroth order component (which we indicate with ``A'') we integrate both sides of it over the whole sphere and we obtain
\begin{equation}
  \label{eq:phiA-VA}
  \begin{cases}
  \partial_t \varphi_A &= \frac{\mathrm{i}}{2} \nabla_r^2\varphi_A - \mathrm{i}  \int \mathrm{d}\Omega [U \psi]\\
  \nabla_r^2V_A &=  p_A + \int \mathrm{d}\Omega [\psi \psi^*]
  \end{cases}\,,
\end{equation}
where $d\Omega = \sin \theta d\phi d\theta$ and $\nabla^2_r \equiv \frac{\partial^2}{\partial r^2} + \frac{2}{r} \frac{\partial}{\partial r}$. To obtain the equations corresponding to a general $(\ell, m)$ mode we integrate each side of the equations multiplied by the corresponding spherical harmonic function and we obtain
\begin{equation}
  \label{eq:phiLM-VLM}
  \begin{cases}
  \partial_t \varphi_{\ell,m} = \frac{\mathrm{i}}{2} \left( \nabla_r^2 - \frac{\ell(\ell +1)}{r^2} \right) \varphi_{\ell,m} - \mathrm{i}  \int \mathrm{d}\Omega [U \psi] Y^*_{\ell,m}\\
  \left( \nabla_r^2 - \frac{\ell(\ell +1)}{r^2} \right) V_{\ell,m} =  p_{\ell,m} + \int \mathrm{d}\Omega [\psi \psi^*]Y^*_{\ell,m}
  \end{cases}\,.
\end{equation}

%%%%%%%%%%%%%%%%%%%%%%%%%%%%%%%%%%%%%%%%%%%%%%
\section{Running the simulation}
\label{sec:evolving-the-system}
%%%%%%%%%%%%%%%%%%%%%%%%%%%%%%%%%%%%%%%%%%%%%%
In order to study the evolution of our two body system we evolve the SP-equations for each $(\ell, m)$ mode to account for the scalar field structure, along with the equations of motion of the point particle, which are given by Newton's laws of motion. This is accomplished with the following algorithm:
\begin{enumerate}
\item Given the initial conditions, calculate the gravitational potential of the ``scalar field + particle'' system using a sparse matrix solver included in SciPy \cite{scipy};
\item From $t=0$ to $t=\Delta t$, use the iterated Crank-Nicolson method \cite{Teukolsky:1999rm} to advance the scalar field and the Euler's method \cite{hairer2008solving} to advance the position and velocity of the particle;
\item Calculate the gravitational potential  of the ``scalar field + particle'' system using a sparse matrix solver;
\item From $t=\Delta t$ to $t= 2 \Delta t$, use the iterated Crank-Nicolson method to advance the scalar field and the two-step Adams-Bashforth \cite{hairer2008solving} method to advance the position and velocity of the particle;
\item Repeat the former and the latter steps until $t=700$.
\end{enumerate}
%
%
%%%%%%%%%%%%%%%%%%%%%%%%%%%%%%%%%%%
\subsection{The initial conditions\label{subsec:initial-conditions-particle}}
%%%%%%%%%%%%%%%%%%%%%%%%%%%%%%%%%%%
At $t=0$, the scalar field is given by the groundstate, 0-node, stable configuration that was calculated in Section~\ref{subsec:SP-stationary}. The point particle's degrees of freedom are its initial position and velocity. We performed several simulations to study the influence of varying these parameters on the evolution of the system. In each simulation, the initial condition for the field and the gravitational potential are
\beq
\varphi_A(t=0) &=& f_E(r)\,, \quad V_A(t=0) = U_E(r)\,,\\
\varphi_{\ell,m}(t=0)&=& 0\,,\quad V_{\ell,m}(t=0)=0\,,\nonumber
\eeq
where $f_E$ and $U_E$ are given by the 0-node solutions of Eq.~\eqref{eq:stationary-solutions-equation}.
\begin{figure}
\centering
\includegraphics[width=0.75\columnwidth]{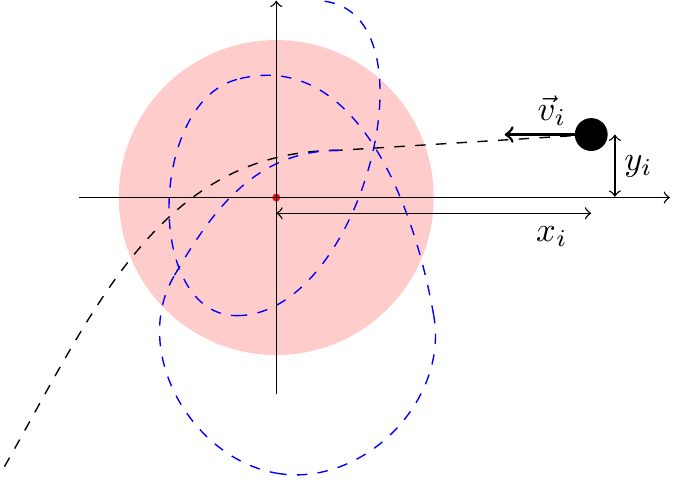}
\caption{Pictorial description of the different outcomes of throwing a pointlike particle at a scalar, self-gravitating structure. We expect that the particle with initial conditions $\vec{r}_i = (x_i, y_i)$, $\vec{v}_i = (-v_i,0)$, being  thrown towards the center of the oscillaton, either scatters (schematically represented in dashed black) or stays in a bounded orbit (in dashed blue).}
\label{fig:scheme}
\end{figure}
The point particle will be thrown at the scalar self-gravitating structure, a setup characterized by (see Fig.~\ref{fig:scheme}),
\begin{enumerate}
\item the impact parameter $y_i$,
\item the mass $M_P$ of the particle being thrown,
\item the velocity $v_i$ with which the particle is thrown.
\end{enumerate}
To write the initial conditions of this motion, we consider the plane that contains the position and velocity vectors of the particle (for all purposes it can be the $\theta = \pi/2$ plane). In this plane, we put the center of coordinates in the center of the scalar field structure, we use $(x_i,y_i)$ to indicate the initial position of the perturbing particle and $(-v_i,0)$ to indicate its initial velocity. Then, we can obtain the initial conditions in polar coordinates of the plane (see Fig.~\ref{fig:scheme}):
\begin{align}
r_i &= \sqrt{x_i^2 + y_i^2}, \quad \phi_i = \arctan\left(\frac{y_i}{x_i}\right)\,,\\
\dot{r}_i &= - v_i \cos\phi_i, \quad   \dot{\phi}_i = \frac{v_i}{r_i}\sin\phi_i.
\end{align}

We run 27 ($3 \times 3 \times 3$) simulations, spanning the following set of initial conditions
\begin{align}
  &x_i = 8, \quad y_i \in \{1.0, 3.0, 5.0\}, \\
  v_i &\in \{0.3, 0.5, 0.7\}, \quad M_p \in \{0.1, 0.001, 10^{-5}\}.
\end{align}
To make it easier to refer to each simulation, we assign a unique code to each of them. The first character of the code refers to the mass of the particle -- ``L'', ``M'' or ``S'', i.e., large, medium or small -- for, respectively, $M_p = 0.1, 10^{-3}$ or $10^{-5}$; the remaining characters will indicate explicitly the impact parameter $y_i$ and the initial velocity $v_i$. As an example, the simulation that has as initial conditions $y_i=1.0, v_i= 0.3, M_P=0.1$ is called ``simulation \textbf{L}\textit{Y}1\textit{V}03''.

We also run 9 ($3 \times 3$) simulations in which we evolve the equations of motion of the particle and the SP-system side-by-side without considering the effect of the particle on the scalar field. We will call these the ``control tests'' and we will refer to them with a similar code indicating the impact parameter and the initial velocity. So, the ``control test'' that has as initial conditions $y_i=1.0, v_i= 0.3$ is referred to by ``control test \textit{Y}1\textit{V}03''.

%%%%%%%%%%%%%%%%%%%%%%%%%%%%%%%%%%%%%%%%%%%%%%%%%%%%%%%%%%%%%
\subsection{The boundary conditions of the SP-system}
%%%%%%%%%%%%%%%%%%%%%%%%%%%%%%%%%%%%%%%%%%%%%%%%%%%%%%%%%%%%%
Regarding the scalar field components, and in line to the description in Section~\ref{subsec:SP-stationary}, we demand regularity
at the boundaries of all quantities. At the origin, regularity is guaranteed by fixing $\varphi'_A(0) = 0$, $\varphi_{\ell,m}(0) = \varphi'_{\ell,m}(0) = 0$ and $U'(0) = 0$. To treat the boundary condition at infinity, we benefit from the careful analysis made in Ref.~\cite{Guzman:2004wj}. The authors show that in order to treat spatial infinity in this system, one must either put it far enough from the active zone or add a sponge to the simulation so that no reflections at the infinity boundary occur. We choose the first option. All the runs of our code were made with a spatial grid that extends up to $r = 1000$ (in the agreed units). We consider that at the infinity boundary all components of the scalar field and of the potential are zero except the spherical component of the potential, $V_A$. In fact, as the mass of the scalar field structure has to be conserved in the grid, we impose that
%
%\begin{equation}
$V_A(r=1000) = -M_f/1000$,
%\end{equation}
where $M_f$ is given in Eq.~\eqref{eq:properties-equilibrium}.

%%%%%%%%%%%%%%%%%%%%%%%%%%%%%%%%%%%%%%%%%%%%%%%%%%%%%%%%%%%%%
\subsection{Time and space discretization of the system}
%%%%%%%%%%%%%%%%%%%%%%%%%%%%%%%%%%%%%%%%%%%%%%%%%%%%%%%%%%%%%
We tested the code -- with results that can be seen in Appendix~\ref{app:testing-code} -- and based on that we decided to conduct all the simulations  using a spatial grid of $\Delta r = 0.1$ and time step of $\Delta t = 10^{-3}$.

%%%%%%%%%%%%%%%%%%%%%%%%%%%%%%%%%%%%%%%%%%%%%%%%%%%%%%%
\section{Results}
\label{sec:results}
%%%%%%%%%%%%%%%%%%%%%%%%%%%%%%%%%%%%%%%%%%%%%%%%%%%%%%%

In the simulations of the ``scalar field + particle'' system, the evolution of both components encodes information about the whole system. We will show details regarding the movement of the particle and the struture of the scalar field configuration as a function of time. Particularly, we will analyse how the backreactions of the field affect the movement of the particle and how the non-spherical components of the field evolve.

%%%%%%%%%%%%%%%%%%%%%%%%%%%%%%%%%%%%%%%%%%%%%%
\subsection{General evolution of the field}
%%%%%%%%%%%%%%%%%%%%%%%%%%%%%%%%%%%%%%%%%%%%%%
We verified that in the simulations we ran, the description of all the quantities involved -- the scalar field, the gravitational potential, and the trajectory of the point-particle -- were dominated by the $\ell=0$ and $\ell = 1$ terms of the expansions in Eqs.~\eqref{eq:field-components} and \eqref{eq:potential-components}. Particularly, we verify that for the cases with particle mass given by $M_p = 10^{-3}$ and $M_p = 10^{-5}$ the terms $\ell \geq 2$ are completely negligible, whereas for the case with $M_p = 0.1$, we verify that $\mathrm{max}[|\varphi_{2,m}|/|\varphi_{1,m}|] \sim \mathrm{max}[ |V_{2,m}|/|V_{1,m}|] \sim \mathcal{O}(10^{-2})$, which is the upper bound for any other ratio of the form $\mathrm{max}[|\varphi_{\ell+1,m}|/|\varphi_{\ell,m}|]$ or $ \mathrm{max}[ |V_{\ell+1,m}|/|V_{\ell,m}|]$, for $\ell > 1$ in the $M_p = 0.1$ case. The latter fact is translated in a slight change in the numerical values of some of the quantities that are calculated in what follows. However, since our focus will be on orders of magnitude and not in exact numerical values, we will, for the sake of simplicity and economy in the length of the expressions, use throughout this section a truncated series to describe the meaningful quantities, i.e.,
\begin{align}
  \psi(r,\theta,\phi) & =& \varphi_A(r) + \sum_{m = -1}^1 \varphi_{1, m}(r) Y_{1 m} (\theta,\phi), \label{eq:field-components-trunc}\\
  U(r,\theta,\phi) & =& V_A(r)+ \sum_{m = -1}^1 V_{1, m}(r) Y_{1 m} (\theta,\phi)\label{eq:potential-components-trunc}.
\end{align}
Notice that the term $m=0$ isn't considered in this expansion. This is due to the fact that the orbital plane is taken to be $\theta = \pi/2$ (see Section~\ref{subsec:initial-conditions-particle}), and so the $m=0$ components are identically zero.

%%%%%%%%%%%%%%%%%%%%%%%%%%%%%%%%%%%%%%%%%%%%%%%%
\subsection{Effects on the orbiting particle}
%%%%%%%%%%%%%%%%%%%%%%%%%%%%%%%%%%%%%%%%%%%%%%%%
The set of initial conditions of the orbiting particle (see Section~\ref{subsec:initial-conditions-particle}), gives rise to \textit{bounded and unbounded orbits of the equilibrium scalar field structure}. Technically, an unbounded orbit has energy per unit mass,
\begin{equation}
\varepsilon = \frac{1}{2} \left(\dot{r}^2 + r^2 \dot{\phi}^2\right) - U_E(r)\,,\label{eq:energy}
\end{equation}
larger than zero. In such case the expression
\begin{equation}
\dot{r}^2 = 2\varepsilon - U_{\text{eff}}\,,
\end{equation}
is always non-negative, where $U_{\text{eff}}$ is given by
\begin{equation}
U_{\text{eff}} = \frac{\mathcal{L}}{r^2} - 2 U_E(r)\,,
\end{equation}
for initial angular momentum per unit mass\footnote{This classification is valid for the equilibrium configuration, in which angular momentum is conserved.} $\mathcal{L}$. In our case, however, whenever we say that an orbit is unbounded, we simply mean that the apocenter of the orbit wasn't observed in the grid during the whole simulation time. In this sense, we find that all simulations result in bounded orbits except \textbf{M}\textit{Y}1\textit{V}07, \textbf{S}\textit{Y}1\textit{V}07, \textit{Y}1\textit{V}07, \textbf{M}\textit{Y}3\textit{V}07, \textbf{S}\textit{Y}3\textit{V}07, \textit{Y}3\textit{V}07, \textbf{L}\textit{Y}5\textit{V}07, \textbf{M}\textit{Y}5\textit{V}07, \textbf{S}\textit{Y}5\textit{V}07 and \textit{Y}5\textit{V}07. Notice that all the unbounded orbits correspond to initial velocity $v_i = 0.7$ and that for the cases $y_i = 1.0, 3.0$ the control test is unbounded whereas the corresponding simulation with $M_P = 0.1$ is bounded. These cases show that the reaction of the scalar field to the presence of the massive particle alters significantly its trajectory.
%
% \begin{figure}
%   \centering
%   \includegraphics[width=\columnwidth]{radialC-C71-C7}
%   \caption{Comparing the evolution of the radial coordinate of a particle with mass $M_P = 0.1$ with (C71 case in black) and without (C7* case in blue) backreactions taken into account. One sees that the effect of backreactions changes drastically the movement of the particle, binding it more tightly to the system.}
%   \label{fig:radialC-bound-unbound}
% \end{figure}

%%%%%%%%%%%%%%%%%%%%%%%%%%%
\subsubsection{Friction force}
%%%%%%%%%%%%%%%%%%%%%%%%%%%

The backreaction of the scalar configuration on the motion of the particle can be computed through the calculation of the effective force that appears in the movement of the particle as it travels through the scalar cloud. We call this a ``friction force''.
To calculate it, we are going to compare the acceleration vector in the orbital plane of the simulations with backreactions (indicated by the subscript ``sim'') with the respective ``control tests''\footnote{Remember the classification introduced in subsection~\ref{subsec:initial-conditions-particle}} (indicated by the subscript ``control''), i.e., our friction force is written as
\begin{equation}
\vec{F}_{\text{f}} = \vec{F}_{\text{sim}} - \vec{F}_{\text{control}}\,,
\end{equation}
with $\vec{F}_{\text{control}} = M_p \vec{a}_{\text{control}}$ (same thing for ``sim'' component); the acceleration vector is written as
\begin{equation}
\vec{a} = \left[\frac{d^2 r}{d t^2} - r\left(\frac{d \phi}{dt}\right)^2\right]\hat{r} + \left[r \frac{d^2 \phi}{dt^2} + 2 \frac{d r}{dt}\frac{d \phi}{dt}\right]\hat{\phi}\,.
\end{equation}
So, we will write
\begin{equation}
\left(\frac{\vec{F}_{\text{f}}}{M_P} \equiv\right)\vec{f}_{\text{f}} = \vec{a}_{\text{sim}} - \vec{a}_{\text{control}}\,.
\end{equation}
Another way of quantifying this force is through
%to calculate the quantity\footnote{$\nabla f(r,\phi) = \frac{\partial f}{\partial r} \hat{r} + \frac{1}{r} \frac{\partial f}{\partial \phi} \hat{\phi}$}
%
\begin{equation}
\vec{f}_f = -\nabla U_{\text{sim}}(r,\phi) + \nabla U_{\text{control}}(r),
\end{equation}
where $U$ represents the gravitational potential. The difference between the two should reflect the extra force that appears as a finite mass effect.
The gravitational potential in Eq.~\eqref{eq:potential-components-trunc} can be written as\footnote{We verify that the source terms of the Poisson equation for the $\ell = 1$ components of the gravitational potential -- see Eq.~\eqref{eq:potential-components} --, given by $s_{1,\pm 1} = \int\int (\psi \psi^* + p_{1,\pm 1}Y_{1,\pm 1}) Y_{1,\pm 1}^* \sin\theta \mathrm{d}\theta \mathrm{d}\phi$ where $p_{1,\pm 1}$ comes from the point-particle contribution (see Eqs.~\eqref{eq:SP-scaled-dirac} and \eqref{eq:dirac-delta-particle}), satisfy the following identity $(s_{1,-1} + s_{1,-1}^*)$ = $-(s_{1,1} + s_{1,1}^*)$ and $(s_{1,-1} - s_{1,-1}^*)$ = $(s_{1,1} - s_{1,1}^*)$, which means that $\mathrm{Re}[s_{1,1}] = -\mathrm{Re}[s_{1,-1}]$ and $\mathrm{Im}[s_{1,1}] = \mathrm{Im}[s_{1,-1}]$. This relation between the real and imaginary parts of the source terms allows us to write $V_{1,-1} = R(r) + \mathrm{i} I(r)$ and $V_{1,1} = - V_{1,-1}^*$.\label{footnote:explaining-source-terms}}
\begin{equation}
U = V_A(r) + 2 \sqrt{\frac{3}{8\pi}} \sin\theta \Big(R(r)\cos\phi + I(r)\sin\phi\Big),
\end{equation}
where $R\equiv\mathrm{Re}[V_{1,-1}]$ and  $I\equiv\mathrm{Im}[V_{1,-1}]$. In the plane of motion, which we consider to be $\theta = \pi/2$, we can write $\vec{f}_f$ as 
\begin{align}
  \vec{f}_f &= -\bigg[\frac{\partial V_{A\text{sim}}}{\partial r} - \frac{\partial V_{A\text{control}}}{\partial r}+ \nonumber\\
            &+ 2 \sqrt{\frac{3}{8\pi}}\bigg(\frac{\partial R(r)}{\partial r}\cos\phi + \frac{\partial I(r)}{\partial r}\sin\phi\bigg)\bigg] \hat{r} - \nonumber\\
            &- \frac{2}{r}\bigg[\sqrt{\frac{3}{8\pi}} \Big(- R(r)\sin\phi + I(r)\cos\phi\Big)\bigg] \hat{\phi}.
\end{align}

We will calculate the quantity
\begin{equation}
  \left\langle \vec{f}_{\text{f}}  \right\rangle = \frac{1}{t_{\text{out}} - t_{\text{in}}} \int_{t_{\text{in}}}^{t_{\text{out}}} \vec{f}_{\text{f}} \text{ d}t,
\end{equation}
where $t_{\text{in}}, t_{\text{out}}$ are, respectively, the time in which the particle penetrated and left the scalar field structure for the first time.
In order to understand the influence of each of the parameters on the movement of the particle, we run tests in groups of simulations where only one of the parameters is changing:
\begin{enumerate}
\item same mass and initial velocity: MV1 = \{\textbf{L}\textit{Y}1\textit{V}07, \textbf{L}\textit{Y}3\textit{V}07, \textbf{L}\textit{Y}5\textit{V}07\}, MV2 = \{\textbf{M}\textit{Y}1\textit{V}05, \textbf{M}\textit{Y}3\textit{V}05, \textbf{M}\textit{Y}5\textit{V}05\};
\item same mass and impact parameter: MI1 = \{\textbf{L}\textit{Y}1\textit{V}03, \textbf{L}\textit{Y}1\textit{V}05, \textbf{L}\textit{Y}1\textit{V}07\}, MI2 = \{\textbf{M}\textit{Y}1\textit{V}03, \textbf{M}\textit{Y}1\textit{V}05, \textbf{M}\textit{Y}1\textit{V}07\};
\item same velocity and impact parameter: VI1 = \{\textbf{L}\textit{Y}1\textit{V}07, \textbf{M}\textit{Y}1\textit{V}07, \textbf{S}\textit{Y}1\textit{V}07\}, VI2 = \{\textbf{L}\textit{Y}1\textit{V}05, \textbf{M}\textit{Y}1\textit{V}05, \textbf{S}\textit{Y}1\textit{V}05\}.
\end{enumerate}
\begin{figure}
  \centering
  \includegraphics[width=\columnwidth]{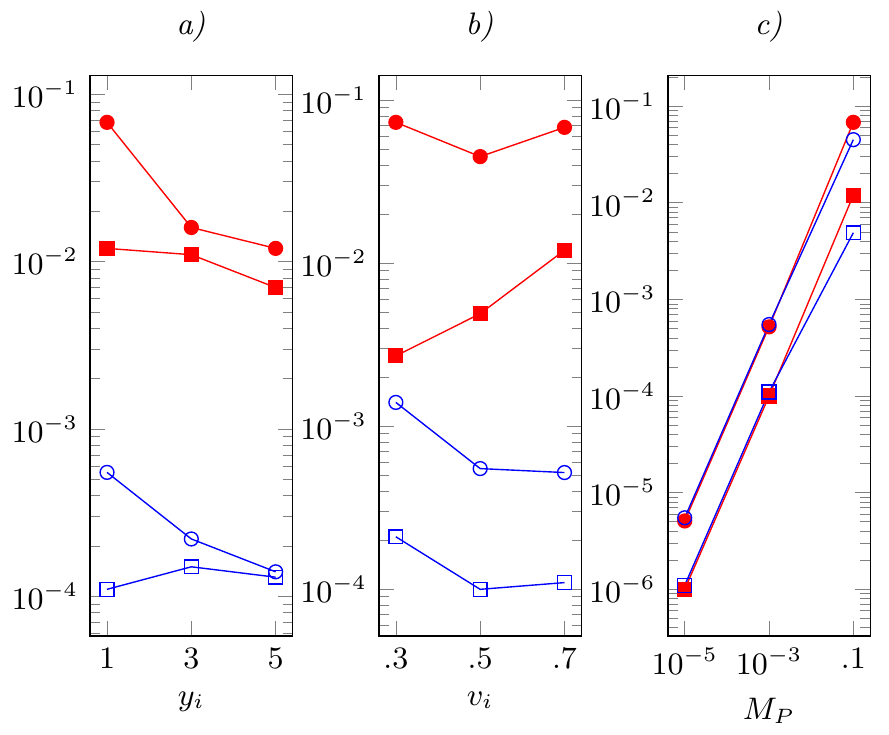}
  \caption{Representing $-\langle f^r_{\text{f}}\rangle$ (by $\bullet$ or $\circ$) and $-\langle f^{\phi}_{\text{f}}\rangle$ (by $\blacksquare$ or $\square$) for different sets of simulations. In panel a), MV1, MV2, i.e. mass and velocity are kept constant; red, filled points represent $M_P = 0.1, v_i = 0.7$ and blue, hollow points represent $M_P = 0.001, v_i = 0.5$. In panel b) MI1, MI2 i.e. mass and impact parameter are kept constant; red, filled points represent $M_P = 0.1, y_i = 1.0$ and blue, hollow points represent $M_P = 0.001, y_i = 1.0$. In panel c) VI1, VI2 i.e. velocity and impact parameter are kept constant; red, filled points represent $v_i = 0.7, y_i=1.0$ and blue, hollow points represent $v_i = 0.5, y_i=1.0$}
  \label{fig:friction-various-tests}
\end{figure}
One can make two comments on the results of this exercise. The first is that the mass is the factor that influences the most the magnitude of the friction force\footnote{This is not surprising in our context given that the range of variation of the value of the mass is much bigger than the one of the other two parameters. We can say that we are conditioned by the size of the scalar field structure. If the impact parameter and/or the velocity are too big or too small the simulations won't work properly either because the particle wouldn't spend enough time close to the scalar field structure or because it would pass too far from it in such a way that the interaction wouldn't produce measurable effects.}. As can be seen in Fig.~\ref{fig:friction-various-tests}, the variation of the initial velocity $v_i$ and the impact parameter $y_i$ does not affect significantly the order of magnitude of the friction force, whereas one can see a systematic and clear variation of this value as one changes the value of the mass of the particle. Specifically, one verifies that
\begin{equation}
  \label{eq:force-vs-mass}
 - \left\langle f^{r}_{\text{f}}  \right\rangle \sim \alpha_r M_P, \quad -\left\langle f^{\phi}_{\text{f}}  \right\rangle \sim \alpha_{\phi} M_P,
\end{equation}
where $\alpha_r$ depends on the initial velocity and the impact parameter, while $\alpha_{\phi}$ almost doesn't depend on those parameters, presenting a value of the order $10^{-1}$ in all instances. The dependence of $\alpha_r$ in the $y_i$ and $v_i$ parameters is asymmetrical: the variation of the initial velocity doesn't affect significantly its value -- the order of magnitude does not change -- whereas the bigger the impact parameter the smaller is the order of magnitude of the coefficient. In fact, while for cases in which $y_i = 1, 3$ we verify $\alpha_r \sim \mathcal{O}(1)$ for $y_i = 5$ we observe $\alpha_r \sim \mathcal{O}(10^{-1})$. This comes as no surprise since the bigger the impact parameter the farther from the center of the scalar structure the particle will pass which means that the particle crosses regions where the scalar field is more and more diluted, decreasing the value of the friction force. Moreover, since the value we are calculating corresponds to a force per unit mass, which by the relations of Eq.~\eqref{eq:force-vs-mass} is proportional to the mass of the incoming particle, we conclude that the total force $\vec{F}$ scales as the square of the incoming particle $M_p$. This is not a new result (see, for instance \cite{Ostriker:1998fa}), but provides a connection between our study and the study of the drag force in self-interacting media.

The second comment has to do with the direction of the average force. We obtain that the average friction force has negative components in both planar directions.

%%%%%%%%%%%%%%%%%%%%%%%%%%%%%%%%%%%%%%%%%%%%
\subsubsection{Loss of angular momentum}
%%%%%%%%%%%%%%%%%%%%%%%%%%%%%%%%%%%%%%%%%%%

\begin{figure}
  \centering
  \includegraphics[width=\columnwidth]{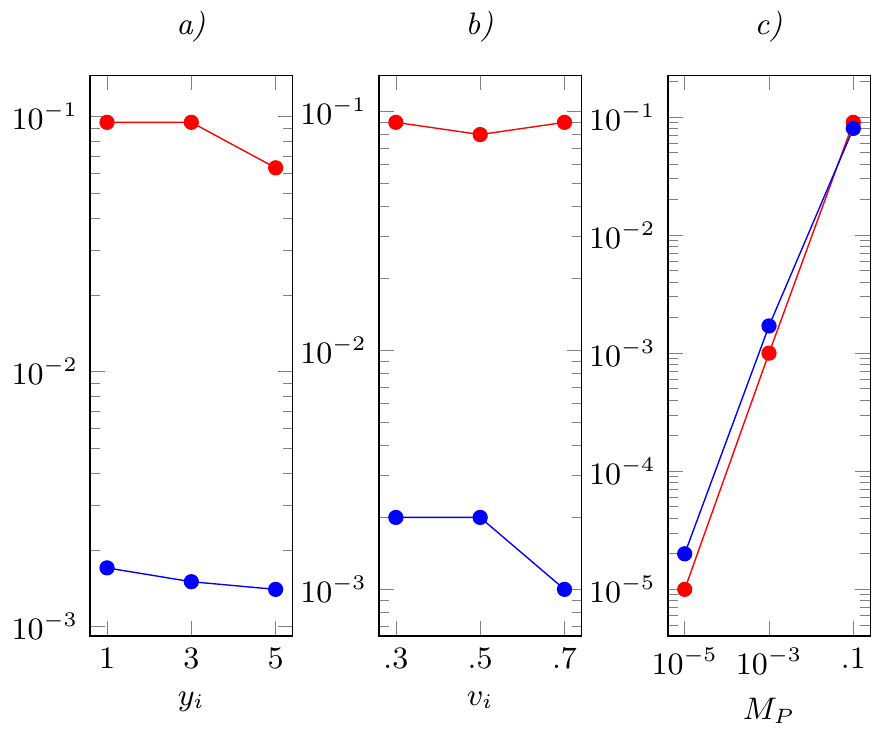}
  \caption{Representing $-\Delta \mathcal{L}$ (see Eq.~\eqref{eq:angularM-difference}) for different sets of simulations. In panel a), MV1, MV2, i.e. mass and velocity are kept constant; red points represent $M_P = 0.1, v_i = 0.7$ and blue points represent $M_P = 0.001, v_i = 0.5$. In panel b) MI1, MI2 i.e. mass and impact parameter are kept constant; red points represent $M_P = 0.1, y_i = 1.0$ and blue points represent $M_P = 0.001, y_i = 1.0$. In panel c) VI1, VI2 i.e. velocity and impact parameter are kept constant; red points represent $v_i = 0.7, y_i=1.0$ and blue points represent $v_i = 0.5, y_i=1.0$}
  \label{fig:angular-momentum-loss}
\end{figure}

Another way to describe the effect of the interaction between the orbiting particle and the scalar field is to study the angular momentum of the former. We don't use the energy because the energy depends on the gravitational potential which in our scenario is dynamical and so it doesn't provide a good measure to characterize the movement of the orbiting particle. The angular momentum, however, is a good measure in the sense that it depends only on kinematic variables. Having said that, we are going to study the quantity
\begin{equation}
  \label{eq:angularM-difference}
  \Delta \mathcal{L} = \frac{\mathcal{L}_{\text{out}} - \mathcal{L}_{\text{in}}}{\mathcal{L}_{\text{in}}}
\end{equation}
where $\mathcal{L}_{\text{out}}$, $\mathcal{L}_{\text{in}}$ represent the angular momentum per unit mass of the orbiting particle when it leaves and when it enters the scalar field structure, respectively. We show in Fig.~\ref{fig:angular-momentum-loss}, the angular momentum per unit mass a function of time. From the picture, we can see that, as expected, the loss of angular momentum is mainly affected by the mass of the orbiting particle. Particularly, the following relation can be found
\begin{equation}
  \Delta \mathcal{L} = \sigma M_P
\end{equation}
with the proportionality factor, $\sigma$, varying slightly with the initial velocity and impact factor being, however, always of order $10^{-1}$.

%%%%%%%%%%%%%%%%%%%%%%%%%%%%%%%%%%%%%%%%%%%%%%%%%%%%%%%%%%%%%
\subsection{Changes in the density distribution of the field}
%%%%%%%%%%%%%%%%%%%%%%%%%%%%%%%%%%%%%%%%%%%%%%%%%%%%%%%%%%%%%
The appearance of the friction force and the loss of angular momentum are related to the dynamical reaction of the scalar field to the presence of the incoming particle. In this respect, it is verified that the scalar field structure develops non-spherical over-densities that are time dependent. In order to appreciate this behavior, we will isolate the different components of the scalar field density. Using Eqs.~\eqref{eq:field-components-trunc} and \eqref{eq:potential-components-trunc} we can write the quantity $\rho_f = (4\pi)^{-1} \psi \psi^*$ (see Eq.~\eqref{eq:stationary-mass}) as
\begin{equation}
  \rho_f = \frac{1}{4\pi} \left(\rho_A + \rho_{1,-1} Y_{1-1} + \rho_{1,1} Y_{11}\right),
\end{equation}
where
\beq
  \rho_A &=& \varphi_A\varphi_A^* + \frac{3 \sin^2\theta}{8 \pi}\left( \varphi_{1,-1}\varphi_{1,-1}^* +  \varphi_{1,1}\varphi_{1,1}^*\right),\\
  \rho_{1,-1} &=& \varphi_{1,-1}\varphi_A^* - \varphi_A\varphi_{1,1}^*\,,\\
  \rho_{1,1} &=& \varphi_{1,1}\varphi_A^* - \varphi_A\varphi_{1,-1}^*.
\eeq
To simplify the expression for $\rho_f$, we observe that (see footnote~\ref{footnote:explaining-source-terms})
\begin{equation}
\mathrm{Re}[\rho_{1,-1}] = - \mathrm{Re}[\rho_{1,1}], \quad  \mathrm{Im}[\rho_{1,-1}] = \mathrm{Im}[\rho_{1,1}],
\end{equation}
which allows us to write, without loss of generality,
\be
\label{eq:A-B-rho-DEF}
\rho_{1,-1}(t,r)= A(t,r) + \mathrm{i}B(t,r)\,,\,\, \rho_{1,1}(t,r) =-\rho_{1,-1}^*(t,r)\,,
\ee
and with that we can rewrite $\rho_f$ as 
\begin{equation}
  \label{eq:density-with-AB}
%  \rho_f = \frac{1}{4\pi} \left(\rho_A + \sqrt{\frac{3}{2\pi}} \sin\theta \left[A(t,r) \cos\phi + B(t,r) \sin \phi\right]\right).
  \rho_f = \frac{1}{4\pi} \left(\rho_A + \sqrt{\frac{3}{2\pi}} \left[A(t,r) \cos\phi + B(t,r) \sin \phi\right]\right),
\end{equation}
where we fixed the value $\theta = \pi/2$ for the orbital plane.

%%%%%%%%%%%%%%%%%%%%%%%%%%%%%%%%%%%%%%%%%%%%%%%%%%%%%%%%%%%%%
\subsubsection{Time dependence of the non-spherical density}
\begin{figure}
  \centering
  \includegraphics[width=0.9\columnwidth]{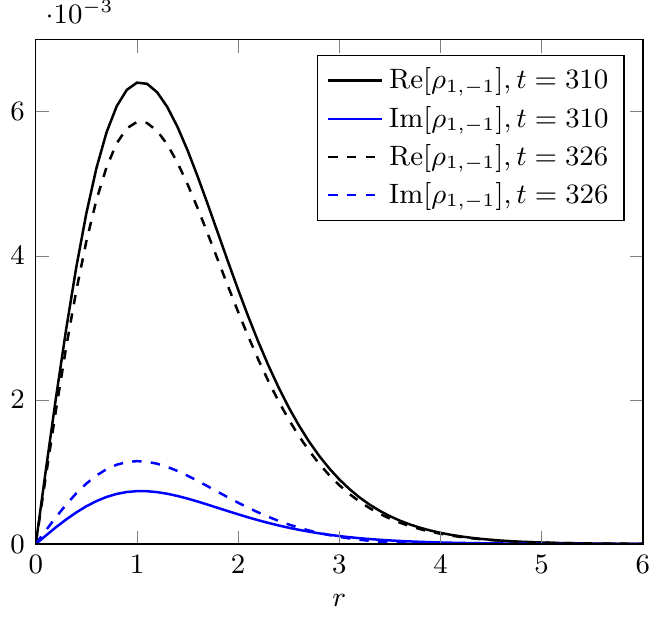}
  \caption{Representing the non-spherical components ($A(t,r)$ = $\mathrm{Re}[\rho_{1,-1}]$ and $B(t,r)$ = $\mathrm{Im}[\rho_{1,-1}]$) of the scalar field density obtained from simulation \textbf{M}\textit{Y}5\textit{V}03 in two different instants of time. In these two instants of time, we see that as the maximum value of the real component decreases, the imaginary component one increases.}
  \label{fig:halo_profile_t326}
\end{figure}

\begin{figure*}
\subfloat{\includegraphics[width = 2\columnwidth]{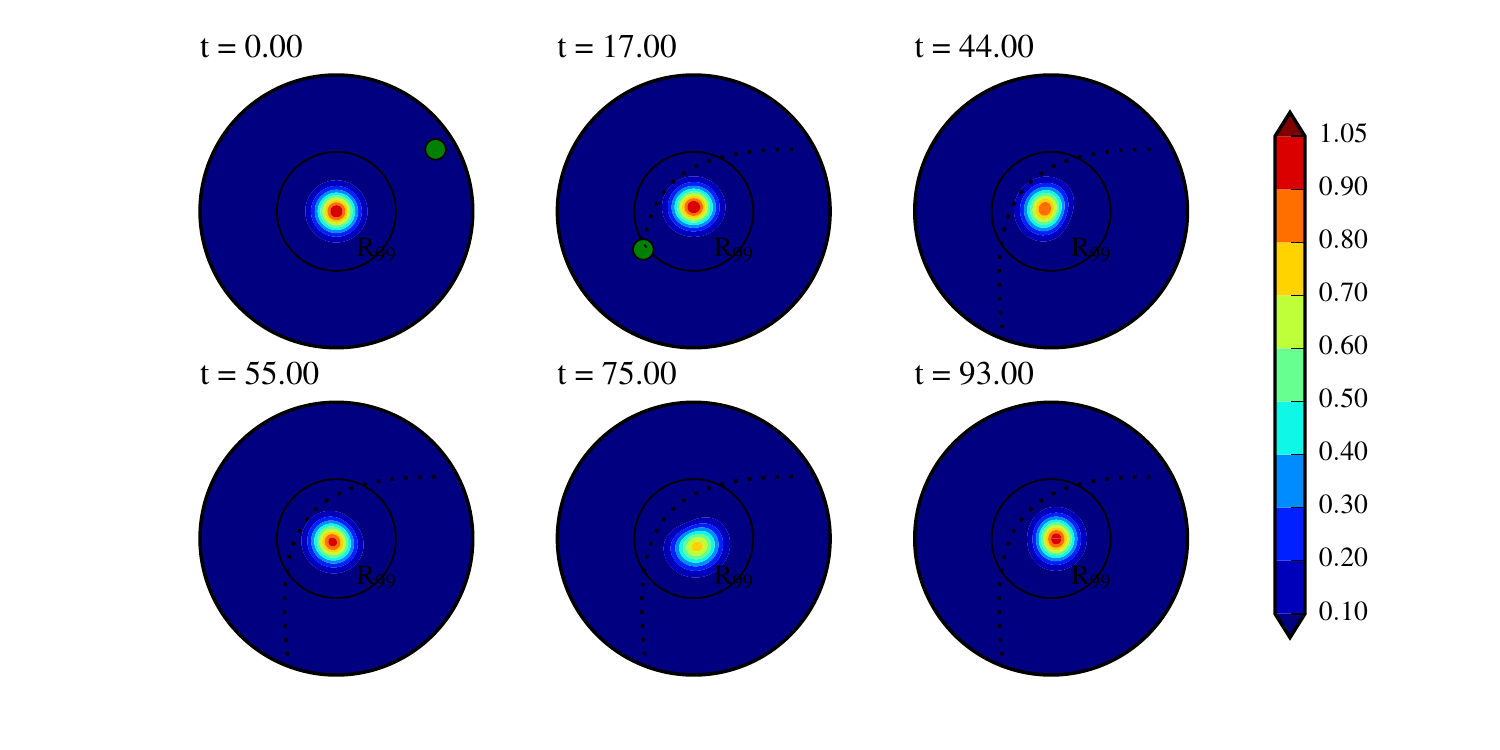}}\\
%\subfloat{\includegraphics[width = 2\columnwidth]{radialC-PLOTS-with-POINTS-D71}}
\caption{We represent the contour plot of the density of the field as in Eq.~\eqref{eq:density-with-AB} for different instants of time using the simulation \textbf{L}\textit{Y}5\textit{V}07. The pointlike particle is shown as a green circle and $R_{99}$ represents the radius of the stable scalar field configuration (see Eq.~\eqref{eq:properties-equilibrium}). It is clear that after the passing of the point particle, the scalar field density develops a rotating movement around the origin of coordinates.}
\label{fig:crazy_big_figure}
\end{figure*}

%%%%%%%%%%%%%%%%%%%%%%%%%%%%%%%%%%%%%%%%%%%%%%%%%%%%%%%%%%%%%
The time dependence of the functions $A(t,r)$ and $B(t,r)$ (see Eq.~\eqref{eq:A-B-rho-DEF}) is illustrated in Fig.~\ref{fig:halo_profile_t326}. In the figure, we see that the profile of the non-spherical components of the density evolves with time, a behavior that, combined with the angular dependence conveyed by the sinusoidal functions (see Eq.~\eqref{eq:density-with-AB}), will result in rotating and oscillating non-spherical component of the density of the scalar field.  A dramatic example of such behavior can be appreciated in Fig.~\ref{fig:crazy_big_figure} in which it is displayed the value of the density function in the plane $\theta = \pi/2$ for the simulation \textbf{L}\textit{Y}5\textit{V}07. In this particular simulation, the incoming particle is scattered by the scalar field structure and moves past it. However, the short time in which the particle is close to the center of the scalar field structure is enough to give rise to a rotating over-density, as can be seen in the last contour plot presented in the respective figure. This example is simple to represent, however, when the particle stays in a bounded orbit, the non-spherical over-densities have a less organized behavior. In this case, there are two competing effects: on the one hand the scalar field structure is dictating the evolution of the non-spherical densities through the SP system and on the other hand the movement of the bounded particle introduces an oscillating ``forcing term'' on top of that.

In order to illustrate this behavior we will consider the position of the center of mass of the scalar field configuration, which, as shown below, is directly related to the functions $A$ and $B$ (see Eq.~\eqref{eq:A-B-rho-DEF}).
The position of the center of mass of the scalar field configuration $\vec{r}_{CM} = (x_{CM},y_{CM},z_{CM})$ is given by
\begin{equation}
  \vec{r}_{CM} = \frac{\int \rho_f(\vec{r}) \vec{r} \mathrm{d}^3r}{ \int \rho_f(\vec{r}) \mathrm{d}^3r},
\end{equation}
where $\rho_f(\vec{r})$ is given by Eq.~\eqref{eq:density-with-AB}. Then, the denominator is written as
\begin{align}
  M_f &= \int \rho_f(\vec{r}) \mathrm{d}^3r = \frac{1}{4 \pi} \Bigg( 4 \pi \int \varphi_A^* \varphi_A r^2 \mathrm{d}r + \nonumber\\
  &+ W_1 \left[ - \int \varphi_{1,-1}^* \varphi_{1,-1} r^2 \mathrm{d}r  - \int \varphi_{1,1}^* \varphi_{1,1} r^2 \mathrm{d}r \right] \Bigg),
\end{align}
and since $W_1 = \iint Y_{1-1}Y_{1,1}\sin\theta \mathrm{d}\theta \mathrm{d}\phi = -1$
we obtain that
\begin{align}
  M_f &= \int \varphi_A^* \varphi_A r^2 \mathrm{d}r + \nonumber\\
  &+ \frac{1}{4 \pi} \left( \int \varphi_{1,-1}^* \varphi_{1,-1} r^2 \mathrm{d}r  + \int \varphi_{1,1}^* \varphi_{1,1} r^2 \mathrm{d}r \right).
\end{align}

In all our simulations the movement is planar, so it suffices to calculate the $(x,y)$ coordinates of the center of mass (the origin of the coordinates is at the center of the initial configuration of the scalar field). We obtain that
\begin{equation}
  \label{eq:x-cm}
  x_{\mathrm{CM}} M_f = \frac{1}{3} \sqrt{\frac{3}{2\pi}} \int r^3 A(t,r) \mathrm{d}r,
\end{equation}
and
\begin{equation}
  \label{eq:y-cm}
  y_{\mathrm{CM}} M_f = \frac{1}{3} \sqrt{\frac{3}{2\pi}} \int r^3 B(t,r) \mathrm{d}r.
\end{equation}

\begin{figure}
  \centering
  \includegraphics[width=\columnwidth]{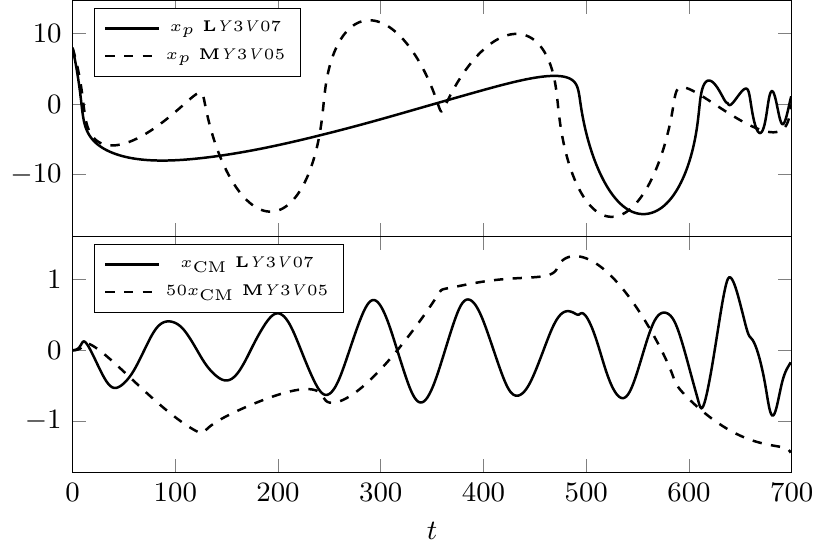}
  \caption{Representing the evolution in time of the $x$-coordinate of the center of mass of the scalar field structure $(x_{\mathrm{CM}})$ and of the orbiting particle $(x_P)$ for simulations \textbf{L}\textit{Y}3\textit{V}07 and \textbf{M}\textit{Y}3\textit{V}05. Both simulations represent bounded orbits and they differ only in the mass of the orbiting particle. Notice that the bigger the mass of the particle, the bigger the value of $x_{\mathrm{CM}}$ and the more oscillations it presents.}
  \label{fig:CM-plots-b}
\end{figure}

\begin{figure}
  \centering
  \includegraphics[width=\columnwidth]{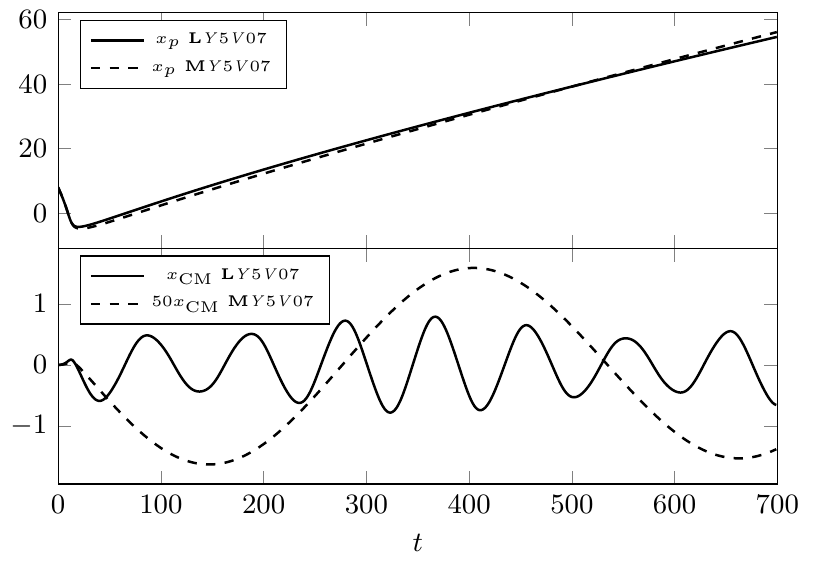}
  \caption{Representing the evolution in time of the $x$-coordinate of the center of mass of the scalar field structure $(x_{\mathrm{CM}})$ and of the orbiting particle $(x_P)$ for simulations \textbf{L}\textit{Y}5\textit{V}07 and \textbf{M}\textit{Y}5\textit{V}07. Both simulations represent unbounded orbits and they differ only in the mass of the orbiting particle. Similarly to the bounded orbit case, the bigger the mass of the particle, the bigger the value of $x_{\mathrm{CM}}$ and the more it oscillates, but in this case, the particle is unbounded. The center-of-mass of the scalar field structure keeps moving even when the particle goes away.}
  \label{fig:CM-plots-u}
\end{figure}

We verify that the center of mass of the scalar configuration oscillates around its initial position -- $x_{\mathrm{CM}}=y_{\mathrm{CM}}=0$ -- and the magnitude of the oscillation depends mainly on the mass of the particle. As we can see in Figs.~\ref{fig:CM-plots-b} and \ref{fig:CM-plots-u}, bounded orbits will produce less organized oscillations of the coordinates of the center of mass whereas unbounded orbits create a more organized, regular pattern. Moreover, independently of the other parameters, one verifies that $\mathcal{O}(x_{\mathrm{CM}}) \sim 10^{-1}$ for $M_P\sim 10^{-1}$, $\mathcal{O}(x_{\mathrm{CM}}) \sim 10^{-2}$ for $M_P\sim 10^{-3}$ and $\mathcal{O}(x_{\mathrm{CM}}) \sim 10^{-4}$ for $M_P\sim 10^{-5}$; the same relations hold for $y_{\mathrm{CM}}$.


\begin{figure}
  \centering
  \includegraphics[width=\columnwidth]{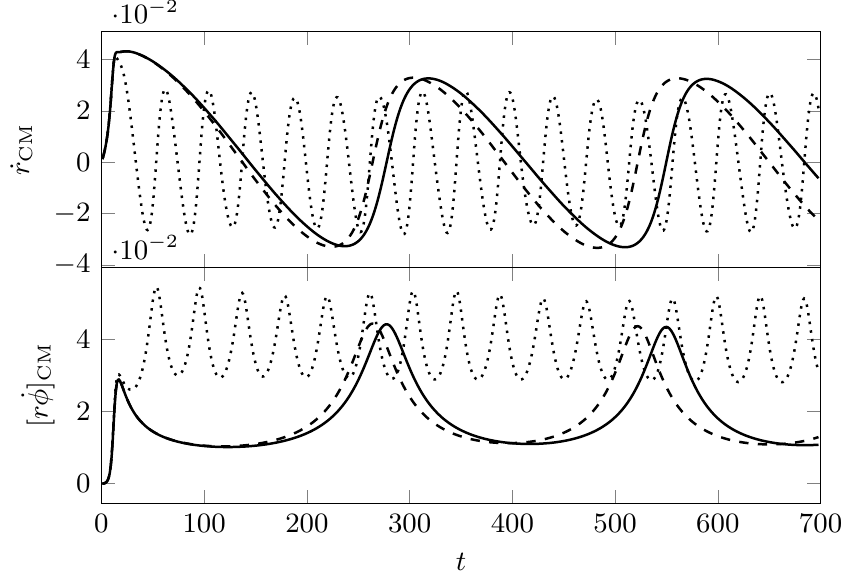}
  \caption{Representing the evolution in time of the radial and angular velocities of the center-of-mass of the scalar field structures -- Eqs.~\eqref{eq:x-cm} and \eqref{eq:y-cm} -- obtained for the simulations \textbf{L}\textit{Y}5\textit{V}07, \textbf{M}\textit{Y}5\textit{V}07 and \textbf{S}\textit{Y}5\textit{V}07. Dotted lines represent the simulation \textbf{L}\textit{Y}5\textit{V}07, dashed lines represent the values of the simulation \textbf{M}\textit{Y}5\textit{V}07 multiplied by $10^2$ and the filled lines represent the CM velocities of simulation \textbf{S}\textit{Y}5\textit{V}07 multiplied by $10^4$. Notice the dependence of the magnitude of the CM velocities on the mass of the incoming particle and the similarities of the frequencies and shape of the evolution of simulations \textbf{M}\textit{Y}5\textit{V}07 and \textbf{S}\textit{Y}5\textit{V}07.}
  \label{fig:angular-radial-velocity-CM}
\end{figure}

As became clear in the study of the friction force and the loss of angular momentum, the mass of the particle is the most important factor determining the change in its dynamics. Taking that into account, we will focus on the characteristics of the movement of the center of mass of the scalar field configuration using the simulations \textbf{S}\textit{Y}5\textit{V}07, \textbf{M}\textit{Y}5\textit{V}07 and \textbf{L}\textit{Y}5\textit{V}07. Using this set of simulations is appropriate for two reasons: 1) all the orbits are unbounded, which allows a clearer and simpler analysis of the dynamical aspects of the center of mass; 2) the simulations differ from each other by the value of the mass of the incoming particle, which is exactly the parameter that influences the most all the details of the dynamics of the system.
We calculate the radial and angular velocity of the center of mass by using the values of $x_{\mathrm{CM}}$ and $y_{\mathrm{CM}}$ that we calculate directly from the simulation files (see Eqs.~\eqref{eq:x-cm} and \eqref{eq:y-cm}). To do it, we use the following expressions
\begin{align}
  r_{\mathrm{CM}} &= \sqrt{x_{\mathrm{CM}}^2 + y_{\mathrm{CM}}^2} \,,\\
  \phi_{\mathrm{CM}} &= \arctan \left(\frac{y_{\mathrm{CM}}}{x_{\mathrm{CM}}}\right)\,, \\
  \dot{r}_{\mathrm{CM}} &=\frac{x_{\mathrm{CM}} \dot{x}_{\mathrm{CM}} + y_{\mathrm{CM}} \dot{y}_{\mathrm{CM}}}{r_{\mathrm{CM}}}\,,\\
  \dot{\phi}_{\mathrm{CM}} &= \frac{\dot{y}_{\mathrm{CM}} x_{\mathrm{CM}} - y_{\mathrm{CM}} \dot{x}_{\mathrm{CM}}}{r^2_{\mathrm{CM}}}.
\end{align}
in which $\dot{x} \equiv dx/dt$. We present the results of these calculations in Fig.~\ref{fig:angular-radial-velocity-CM} and from there two things are evident: 1) the behavior of the simulations with the smaller values of the mass of the particle are very similar, except for the frequency of oscillation and the magnitude; 2) the magnitude of the velocity components scales with the mass of the incoming particle, and we verify that
\begin{equation}
  \mathrm{max} [\dot{r}_{\mathrm{CM}}] = \mathrm{max}[ (r \dot{\phi})_{\mathrm{CM}}] \propto 0.1 M_P.
\end{equation}

%
%
%%%%%%%%%%%%%%%%%%%%%%%%%%%%%%%%%%%%%%%%%%%%%%%%%%%%%%%
\subsubsection{Magnitude of the non-spherical density}
\begin{figure}
  \centering
  \includegraphics[width=\columnwidth]{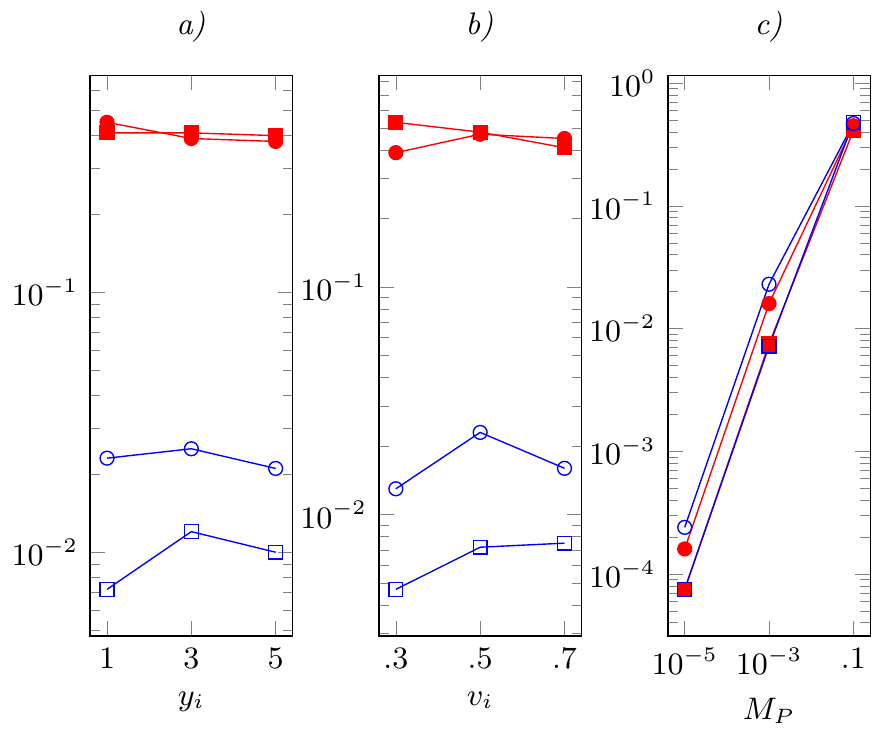}
  \caption{Representing the maximum value in time and space of $\mathrm{Re}[\rho_{1,-1}]$ (by $\bullet$ or $\circ$) and $\mathrm{Im}[\rho_{1,-1}]$ (by $\blacksquare$ or $\square$) for different sets of simulations. In panel a) MV1,MV2, i.e. mass and velocity are kept constant; red, filled points represent $M_P = 0.1, v_i = 0.7$ and blue, hollow points represent $M_P = 0.001, v_i = 0.5$. In panel b) MI1, MI2 i.e. mass and impact parameter are kept constant; red, filled points represent $M_P = 0.1, y_i = 1.0$ and blue, hollow points represent $M_P = 0.001, y_i = 1.0$. In panel c) VI1, VI2 i.e. velocity and impact parameter are kept constant; red, filled points represent $v_i = 0.7, y_i=1.0$ and blue, hollow points represent $v_i = 0.5, y_i=1.0$.}
  \label{fig:density-various-tests}
\end{figure}

%%%%%%%%%%%%%%%%%%%%%%%%%%%%%%%%%%%%%%%%%%%%%%%%%%%%%%%
In Fig.~\ref{fig:halo_profile_t326} it is plotted the profile of the non-spherical components of the scalar field density, namely $A(t,r)$ and $B(t,r)$ for two different moments in time. The magnitude of these non-spherical components depends on the mass of the neighboring particle: the bigger the mass $M_P$, the bigger the magnitude of these componentes, as can be seen in Fig.~\ref{fig:density-various-tests}. There, we represent the maximum value of the magnitude of functions $A$ and $B$ (see Eq.~\eqref{eq:A-B-rho-DEF}) in the different sets of simulations MV1, MV2, MI1, MI2, VI1 and VI2. We see that, again, of the three initial parameters of the incoming particle, the mass has the strongest influence on the magnitude of the non-spherical components of the scalar field density. In fact, similarly to the case of the friction force, one can write
\begin{equation}
  \mathrm{max}\left[A\right] \sim \beta_A M_P, \quad \mathrm{max}\left[B\right] \sim \beta_B  M_P,\label{eq:betaAB}
\end{equation}
i.e., the maximum magnitude of both $A$ and $B$ is directly proportional to the mass of the incoming particle, with the proportionality factors $\beta_{A,B}$ presenting values between 3 and 5 without any correlation with the initial velocity and impact parameter of the particle.
%
%

%%%%%%%%%%%%%%%%%%%%%%%%%%%%%%%%%%%%%%%%%%%%%%%%%%%%%%%
\section{Conclusion}
\label{sec:conclusion}
%%%%%%%%%%%%%%%%%%%%%%%%%%%%%%%%%%%%%%%%%%%%%%%%%%%%%%%
We studied what happens to a stable, low-energy scalar field configuration when a particle-like body passes in its neighborhood. We observe that the mass of such body is the most important factor parametrizing the interaction between the two involved parties. Specifically, we verify that once the incoming particle dives in the scalar field structure, it is affected by a friction-like force that scales as $M_p^2$. This result bridges our calculations with other studies of friction force in self-interacting media. We also studied the effect of this friction force on the loss of angular momentum by the orbiting particle, with results that show the same scaling , i.e., the loss of angular momentum per unit mass scales with $M_p$. We verify the development of non-spherical components of the initially spherical scalar field structure due to the presence of the neighboring particle, their magnitude scaling with the particle's mass $M_p$. Furthermore, we observe that even when the incoming body passes by the scalar field structure, i.e. describing an unbounded orbit, a non-spherical component develops and stays rotating around the center of the scalar strucure. The velocity of rotation of these non-spherical components scales with the mass of the orbiting particle.

The apparent strong connection between the mass of the incoming particle and the effects it leaves on the scalar field structure, may become a good lead in investigations about the history of such structures. In the scenarios in which DM is described by such scalar field structures, the detection of rotating clumps of DM in galaxies may be explained by a primordial encounter between the DM aggregate and a passing massive particle-like body. In such spirit, future work should include a more detailed analysis of the numerical approximations that were considered in the present work, as well as a simulation of a more rich scenario, in which the scalar field structure is orbited by stars as the encounter with the particle-like body happens.

%%%%%%%%%%%%%%%%%%%%%%%%%%%%%%%%%%%%%%%%%%%%%%%%%%%%%%%%%%%%%%%%%%%%%%%%%%%%%%
\begin{acknowledgments}
We thank Miguel Zilhão, Caio Macedo, David Hilditch and, specially, V\'itor Cardoso for their insightful comments and feedback.
We thankfully acknowledge the computer resources, technical expertise and assistance provided by S\'ergio Almeida and Manuel Torrinha at CENTRA/IST. Computations were performed at the cluster ``Baltasar-Sete-S\'ois'', and supported by the MaGRaTh--646597 ERC Consolidator Grant.
The author acknowledges financial support provided by Funda\c{c}\~{a}o para a Ci\^{e}ncia e a Tecnologia Grant number PD/BD/113481/2015 awarded in the framework of the Doctoral Programme IDPASC - Portugal.
\end{acknowledgments}

\appendix

%%%%%%%%%%%%%%%%%%%%%%%%%%%%%%%
\section{Testing the code}
\label{app:testing-code}
%%%%%%%%%%%%%%%%%%%%%%%%%%%%%%

We evolve the two components of our two-body system using different techniques. To solve SP system of equations, of the form of Eqs.~\eqref{eq:phiA-VA} and \eqref{eq:phiLM-VLM}, we use a centered finite difference stencil to write the derivatives. Particularly, at a generic point $u_j = j\Delta u$, we discretize the first derivatives as
\begin{equation}
  \frac{\partial H}{\partial u} = \frac{H_{j+1} - H_{j-1}}{2\Delta u}
\end{equation}
and the second derivatives as
\begin{equation}
  \frac{\partial^2 H}{\partial u^2} = \frac{H_{j+1} -2 H_{j} +  H_{j-1}}{\left(\Delta u\right)^2}
\end{equation}
for a general function $H(u)$, indicating $H(u_j) = H_j$. Having discretized the equations, we apply the iterated Crank Nicolson method with two iterations, following the conclusions of Ref.~\cite{Teukolsky:1999rm}. To solve the equations of motion of the point-particle, which can be cast in the generic form $dv/dt = G(t,v)$, we use Euler's method, with the evolution step given by
\begin{equation}
    \begin{cases}
    v_{n+1} &= v_{n} +  \Delta t G(t_{n},v_{n})\\
    t_{n+1} &= t_{n} + \Delta t
  \end{cases}\,,
\end{equation}
and the two-step Adams-Bashforth method given by
\begin{equation}
  \begin{cases}
    v_{n+2} &= v_{n+1} + \frac{3}{2} \Delta t G(t_{n+1},v_{n+1}) - \frac{1}{2} \Delta t G(t_{n},v_{n})\\
    t_{n+2} &= t_{n+1} + \Delta t
  \end{cases}\,.
\end{equation}

%%%%%%%%%%%%%%%%%%%%%%%%%%%%%%%%%%%%%%%%%%%%%%%%%%%%%%%%%%%%
\subsection{Evolving a stationary scalar field solution}
\begin{figure}
\centering
\includegraphics[width=\columnwidth]{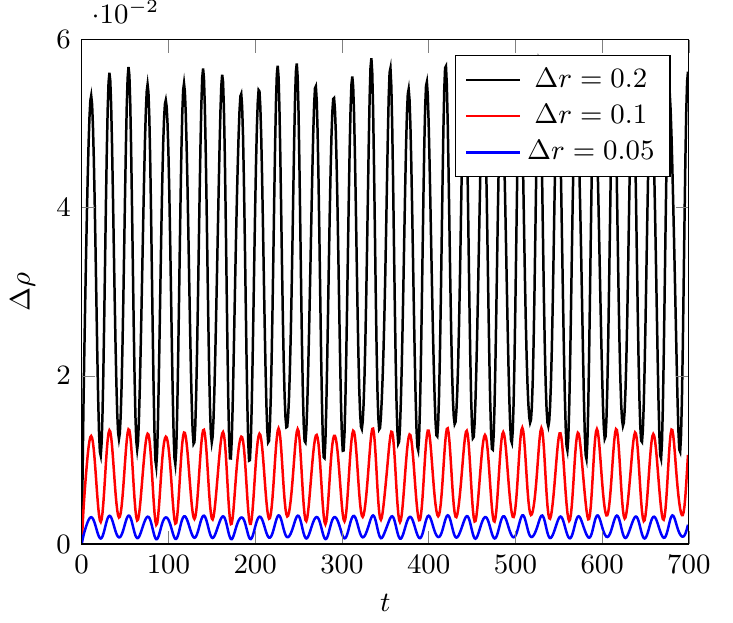}
\caption{Representing the evolution of the quantity $\Delta \rho$ of Eq.~\eqref{eq:delta-rho} using $\Delta t = 10^{-3}$ and with three different grid values. We observe that the maximum value of $\Delta \rho$ in each simulation is related to the grid spacing as $(\Delta r)^2$.}
\label{fig:convergence-field}
\end{figure}

%%%%%%%%%%%%%%%%%%%%%%%%%%%%%%%%%%%%%%%%%%%%%%%%%%%%%%%%%%% 
Using a timestep $\Delta t = 10^{-3}$, we run a test with three different grid spacings - $\Delta r = 0.2, 0.1, 0.05$. To quantify the effect of the grid spacing in the evolution of the field, define
\begin{equation}
\label{eq:delta-rho}
\Delta \rho(t) = \text{max}\left(|\rho_E(r) - \rho(t,r)|\right)\,,
\end{equation}
where $\rho_E(r) = f_E f_E^*$ is the equilibrium density of the scalar field (see Eq.~\eqref{eq:stationary-solutions-equation}) and $ \rho(t,r)$ is the density of the field that is evolved in time using our code. The results of this evolution are shown in Fig.~\ref{fig:convergence-field}.
The test allows us to conclude that with decreasing resolution, the magnitude of the deviations from the initial stationary configuration decreases. Moreover, we obtain that $\mathrm{max}\left[ \Delta \rho \right] \sim \left( \Delta r\right)^2$.

%%%%%%%%%%%%%%%%%%%%%%%%%%%%%%%%%%%%%%%%%%%%%%%%%%%%%%%%%%%%%%%
\subsection{Testing the code evolving the orbiting particle}
%%%%%%%%%%%%%%%%%%%%%%%%%%%%%%%%%%%%%%%%%%%%%%%%%%%%%%%%%%%%%%%
The evolution of the particle will be made with the same time step as the one used for the SP-equations. To correctly describe this evolution, the code has to guarantee the conservation of the energy and angular momentum per unit mass for the control tests. To visualize that conservation, we calculate the following quantities
\begin{equation}
  \label{eq:delta-e-l}
  \Delta \varepsilon = \frac{\varepsilon(t) - \varepsilon(0)}{\varepsilon(0)}, \quad \Delta \mathcal{L} = \frac{\mathcal{L}(t) - \mathcal{L}(0)}{\mathcal{L}(0)},
\end{equation}
where $\varepsilon(0)$ and $\mathcal{L}(0)$ represent the initial energy and angular momentum per unit mass, respectively.
\begin{figure}
  \centering
  \includegraphics[width=\columnwidth]{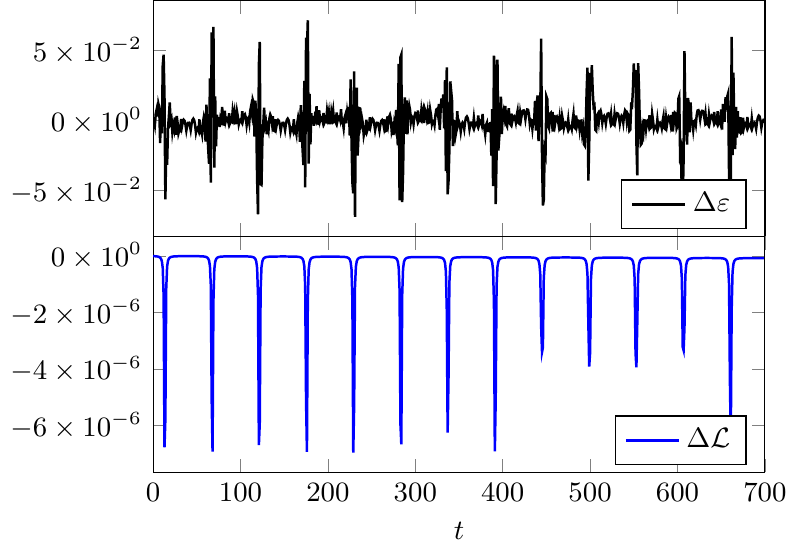}
  \caption{Represening the evolution of the energy and angular momentum per unit mass (see Eq.~\eqref{eq:delta-e-l}) for the simulation \textit{Y}3\textit{V}03. This simulation was run with $\Delta t = 10^{-3}$ and $\Delta r = 0.1$.}
  \label{fig:convergence-particle}
\end{figure}
In Fig.~\ref{fig:convergence-particle} we show the evolution of these quantities for the control test \textit{Y}3\textit{V}03. We observe that both the energy and the angular momentum are conserved, in the worst case, up to the percent level.
%
%
%

%%%%%%%%%%%%%%%%%%%%%%%%%%%%%%%%%%%%%%%%%%%%%%%%%%%%%%%%%%%%%%%%%%%%%%%%%%%%%%%%%%%%%%%%%%%%%%%%
\section{Discretizing the Dirac delta} \label{app:dirac}
%%%%%%%%%%%%%%%%%%%%%%%%%%%%%%%%%%%%%%%%%%%%%%%%%%%%%%%%%%%%%%%%%%%%%%%%%%%%%%%%%%%%%%%%%%%%%%%%
%
In order to describe the perturbing mass orbiting the scalar configuration as a point particle, it is necessary to use the Dirac delta. To describe it in a numerical grid, we follow an approach used in previous works (see Ref.~\cite{Sundararajan:2007jg} and references therein) in which the construction of the discretized version of the Dirac delta is made considering that its defining feature is integrability. This means that we obtain the discretized Dirac delta by studying the expression
\begin{equation}
  \int_{-\infty}^{\infty} f(x) \delta(x-X_0) dx = f(X_0)
\end{equation}
for some ``well-behaved'' (continuous with continuous derivatives) function $f(x)$.
The finite difference version of the previous expression is given by
\begin{equation}
  \label{eq:dirac-finite-difference}
  dx \sum_i f_i \delta_i = f_{i*}
\end{equation}
where $f_i$ and $\delta_i$ represent the values of function $f$ and the Dirac delta, respectively, at the grid point $i$. In our case it suffices to consider that the point $X_0$ is always a grid point such that $X_0/dx = i*$. With this setup, we can say that the only point of the grid in which the Dirac delta takes a non-zero value is precisely the grid point corresponding to $X_0$. This implies that the finite difference formula of Eq.~\eqref{eq:dirac-finite-difference} can be written as
\begin{equation}
  dx f(X_0) \delta_{i*} = f(X_0),
\end{equation}
from where we can read that the Dirac delta has the following finite difference representation
\begin{equation}
  \delta_i = \begin{cases}\frac{1}{dx},& i=i*\\ 0,&i\neq i*\end{cases}.
\end{equation}
This definition agrees, in the respective limits, with the simplest definition for the Dirac delta of \cite{Sundararajan:2007jg}. We decided to use a one-point-only discretized Dirac delta for two reasons: a) it works well; b) given the scales of the problem at hand, we want to reinforce as much as possible the localized nature of the perturbing mass.

%%%%%%%%%%%%%%%%%%%%%%%%%%%%%%%%%%%%%%%%%%%%%%%%%%%%%%%%%%%%%%%%%%%%%%%%%%%%%%%%%%%%%%%%%%%%%%%%
\section{Derivation of the Schrodinger-Poisson system with orbiting particle} \label{app:low-energy}
%%%%%%%%%%%%%%%%%%%%%%%%%%%%%%%%%%%%%%%%%%%%%%%%%%%%%%%%%%%%%%%%%%%%%%%%%%%%%%%%%%%%%%%%%%%%%%%%
In this appendix, we reintroduce the constants $G$, $\hbar$ and $c$ and we will be using the following post-Newtonian metric
\begin{align}
  g_{00} &= -1 + \frac{2}{c^2} U + \mathcal{O}(c^{-4}) \,, \nonumber\\
  g_{0j} &= \mathcal{O}(c^{-3}) \,,\label{eq:post-newton-metric}\\
  g_{jk} &= \left(1  + \frac{2}{c^2}U \right) \delta_{jk} + \mathcal{O}(c^{-4}).\nonumber
\end{align}
\subsection{Poisson equation}
The Einstein tensor, with the metric of Eq.~\eqref{eq:post-newton-metric}, is given by
\begin{align}
  G_{00} &=-\frac{2}{c^2}\nabla^2U + \mathcal{O}(c^{-4})\,,\nonumber\\
  G_{0j} &= \mathcal{O}(c^{-3})\,,\\
  G_{jk} &=  \mathcal{O}(c^{-4}).\nonumber
\end{align}
The energy momentum tensor of the scalar field and the point-particle is written as
\begin{equation}
  T_{\mu\nu} = T_{\mu\nu}^{S} + T_{\mu\nu}^{P}\,,
\end{equation}
where
\begin{equation}
  T_{\mu\nu}^{S} = \frac{1}{2} \Big[ \Phi_{,\mu} \Phi^*_{,\nu} + \Phi_{,\nu} \Phi^*_{,\mu} - g_{\mu\nu} \Big(\Phi^{,\sigma}\Phi^*_{,\sigma} + \frac{m_S^2 c^2}{\hbar^2} |\Phi|^2\Big)\Big]\,,
\end{equation}
corresponds to the energy-momentum tensor of the scalar field, and\footnote{This result is obtained after integrating in time the expression of Eq.~\eqref{eq:em-point-particle-integral}.}
\begin{equation}
  T_{\mu\nu}^P = \frac{g_{\mu \alpha} g_{\nu \beta}}{\sqrt{-g}} M_P \frac{dx_p^{\alpha}}{dt} \frac{dx_p^{\beta}}{dt} \left(\frac{dt}{d\tau}\right) \delta^{(3)}(x - x_p(t(\tau)))\,,
\end{equation}
where $g$ is the determinant of the metric, is the energy-momentum tensor due to a point particle. Using these definitions, one can write
\begin{align}
  \frac{8 \pi G}{c^4} T_{00}^S &= 8\pi G \left(\frac{m_S^2}{2 c^2 \hbar^2}|\Phi|^2  + \frac{1}{2}(\partial_0\Phi)(\partial_0\Phi^*)\frac{1}{c^4} + \mathcal{O}(c^{-4})\right)\,,\nonumber \\
  \frac{8 \pi G}{c^4} T_{0j}^S &= \mathcal{O}(c^{-5})\,,\\
  \frac{8 \pi G}{c^4} T_{jk}^S &= 8\pi G \left(-\frac{m_S^2}{2 c^2 \hbar^2}|\Phi|^2 + \frac{1}{2}(\partial_0\Phi)(\partial_0\Phi^*)\frac{1}{c^4}  + \mathcal{O}(c^{-4})\right)\nonumber\,,
\end{align}
where we separate the term of order $c^{-4}$ from the others because this will be important to define the low-energy regime and where $\partial_0 \equiv \partial/\partial(ct)$ with $t$ indicating the time coordinate.
For the particle, the energy momentum is given by
\begin{align}
  \frac{8 \pi G}{c^4} T_{00}^P &= 8 \pi G \left( \frac{M_P}{c^2} +  \mathcal{O}(c^{-4})\right) \delta^{(3)}(x - x_P)\,,\\
  \frac{8 \pi G}{c^4} T_{0j}^P &= 8 \pi G \left( -\frac{M_P}{c^3} v_j  + \mathcal{O}(c^{-5})\right) \delta^{(3)}(x - x_P)\,,\\
  \frac{8 \pi G}{c^4} T_{jk}^P &= 8 \pi G \left( \frac{M_P}{c^4} v_j v_k + \mathcal{O}(c^{-6})\right) \delta^{(3)}(x - x_P)\,.
\end{align}
Considering that the velocity of the particle is much smaller than the velocity of light, its energy momentum tensor reduces to
\begin{align}
  \frac{8 \pi G}{c^4} T_{00}^P &= 8 \pi G \left( \frac{M_P}{c^2} +  \mathcal{O}(c^{-4})\right) \delta^{(3)}(x - x_P)\,,\nonumber\\
  \frac{8 \pi G}{c^4} T_{0j}^P &=  \mathcal{O}(c^{-3})\,,\\
  \frac{8 \pi G}{c^4} T_{jk}^P &=  \mathcal{O}(c^{-4})\,,\nonumber
\end{align}
which is a reflection of the fact that the behavior of the particle is dominated by its rest mass.
For the scalar field, the low-energy regime also means that its behavior is dominated by its rest mass. Since a scalar field of mass $m_s$ and momentum $p$ oscillates with a frequency $\omega = E/\hbar$, i.e.
\begin{equation}
  \Phi \sim \exp\left(-\mathrm{i} \frac{E}{\hbar} t \right)\,,
\end{equation}
where $E = \sqrt{p^2 c^2 + m_s^2c^4}$ is the energy of the field\footnote{This expression is valid only for free fields, but the rest of the calculations will be valid as long as, up to first order, $E \sim m_s c^2$, which means that the calculation is also valid for bound states of the scalar field whose energy is dominated by the rest-mass term.}, in a low energy regime we rightly consider that
\begin{equation}
  \Phi(t,\vec{x}) = \exp\left(-\mathrm{i} \frac{m_sc^2}{\hbar} t \right) \psi(t,\vec{x})\,,
\end{equation}
where $\psi(t,\vec{x})$ constains not only the spatial dependence of the field but also a slowly varying dependence in time. Because we know that in the low-energy limit the rest-mass dominates, we apply the condition  $p^2 \ll m_s c^2$ such that
\begin{align}
  E &= \sqrt{p^2 c^2 + m_s^2 c^4} = m_sc^2 \sqrt{1 + \frac{p^2}{m_s^2 c^2}} \sim \nonumber\\
  &\sim m_sc^2 \left( 1 + \frac{p^2}{2 m_s^2 c^2}\right).
\end{align}
which allows one to infer that
\begin{equation}
  \psi(t,\vec{x}) \sim \exp\left(-\mathrm{i} \frac{p^2}{2 m_s \hbar} t \right)\,.
\end{equation}

This form of the scalar field reprensents the low-energy limit and the derivative terms in the energy-momentum tensor can thus be simplified as (remember that $\partial_0 \equiv \partial/\partial(ct)$)
\begin{align}
  \frac{1}{2}(\partial_0\Phi)(\partial_0\Phi^*) =& \frac{1}{2 c^2} \left[ -\mathrm{i} \frac{m_sc^2}{\hbar} \Phi + \exp\left(-\mathrm{i}\frac{m_sc^2 t}{\hbar} \right) \partial_t\psi \right] \times \nonumber\\
                                                 & \left[\mathrm{i} \frac{m_sc^2}{\hbar} \Phi^* + \exp\left(\mathrm{i}\frac{m_sc^2 t}{\hbar} \right) \partial_t\psi \right] = \nonumber\\
  &\frac{1}{2 c^2} \left(\frac{m_s^2 c^4}{\hbar^2} \Phi \Phi^* + \partial_t\psi \partial_t\psi\right).
\end{align}
Plugging this result in the expression for the energy-momentum tensor, we obtain  
\begin{widetext}
\begin{align}
  \frac{8 \pi G}{c^4} T_{00}^S &= 8\pi G \left( \left[\frac{m_s^2}{2 c^2 \hbar^2} + \frac{m_s^2}{2 c^2 \hbar^2} \right]|\Phi|^2  + \frac{\partial_t\psi \partial_t\psi}{2 c^6} + \mathcal{O}(c^{-4})\right)\,,\nonumber\\
  \frac{8 \pi G}{c^4} T_{0j}^S &= \mathcal{O}(c^{-5})\,,\\
  \frac{8 \pi G}{c^4} T_{jk}^S &= 8\pi G \left( \left[\frac{m_s^2}{2 c^2 \hbar^2} - \frac{m_s^2}{2 c^2 \hbar^2} \right]|\Phi|^2  + \frac{\partial_t\psi \partial_t\psi}{2 c^6} + \mathcal{O}(c^{-4})\right).\nonumber
\end{align}
\end{widetext}
Notice that $\partial_t\psi \sim p^2/2m_s$ so that
\begin{equation}
  \frac{\partial_t\psi \partial_t\psi}{2 c^6} \sim \frac{p^4}{8 m_s^2 c^6} \sim \mathcal{O}(c^{-6})\,,
\end{equation}
because, again, we are in the low energy limit. Finally, the energy momentum tensor of the scalar field in the low energy limit is given by
\begin{align}
  \frac{8 \pi G}{c^4} T_{00}^S &= 8\pi G \left(\frac{m_s^2}{c^2 \hbar^2} |\Phi|^2 +  \mathcal{O}(c^{-4})\right)\,,\nonumber\\
  \frac{8 \pi G}{c^4} T_{0j}^S &= \mathcal{O}(c^{-5})\,,\\
  \frac{8 \pi G}{c^4} T_{jk}^S &=  \mathcal{O}(c^{-4}).\nonumber
\end{align}
Having defined the low-energy limit, we can now equate both sides of the Einstein equations
\begin{equation}
  G_{\alpha\beta}  = \frac{8 \pi G}{c^4} T_{\alpha\beta}\,,
\end{equation}
which reduce to its 00-component in the low-energy regime:
\begin{equation}
  \nabla^2 U = - 4\pi G \left( M_P \delta^{(3)}(x - x_P) + \frac{m_S^2}{\hbar^2} |\Phi|^2 \right).
\end{equation}

\subsection{Klein-Gordon equation}
Plugging in the Klein-Gordon equation
\begin{equation}
  \frac{1}{\sqrt{-g}} \left(\sqrt{-g} g^{\mu\nu} \Phi_{,\mu}\right)_{,\nu} - \frac{m_S^2 c^2}{\hbar^2} \Phi = 0
\end{equation}
the post-Newtonian metric of Eq.~\eqref{eq:post-newton-metric} and the scalar field 
\begin{equation}
  \Phi(t,\vec{x}) = \exp\left(-\mathrm{i} \frac{m_Sc^2}{\hbar} t \right) \psi(t,\vec{x})\,,
\end{equation}
we obtain
\begin{equation}
  \mathrm{i} \partial_t \psi + \frac{\hbar}{2m_S} \nabla^2 \psi + \frac{m_S}{\hbar} U \psi + \mathcal{O}(c^{-2}) = 0.
\end{equation}
From here, we see that the dynamics described by the Klein-Gordon equation is dominated by the Schrodinger equation in the low-energy limit.

%%%%%%%%%%%%%%%%%%%%%%%
%%%%%%%%%%%%%%%%%%%%%%%
%%%%%%%%%%%%%%%%%%%%%%%
\bibliography{SPbiblio}
\end{document}